\documentclass[11pt]{article}

\usepackage[final]{acl}

\usepackage{times}
\usepackage{latexsym}

\usepackage{booktabs}
\usepackage{ragged2e}
\usepackage{array}
\usepackage{multirow}
\usepackage[table]{xcolor}
\usepackage{graphicx}
\usepackage{amsmath}
\usepackage{amssymb}
\usepackage{mathtools}
\usepackage{amsthm}
\usepackage[most]{tcolorbox}  
\usepackage[dvipsnames]{xcolor}
\usepackage{xcolor}
\usepackage{algorithm}
\usepackage{algorithmic}

\usepackage{hyperref}      
\usepackage{fontawesome5}  
\usepackage{twemojis}      
\usepackage{cuted} 

\definecolor{darkblue}{rgb}{0, 0, 0.5}
\definecolor{SkillMASColor}{RGB}{235, 245, 255}

\usepackage{tikz}
\usepackage{xcolor}
\usepackage{pifont}
\usetikzlibrary{arrows.meta, positioning}

\newcommand{\good}{\textcolor{green!50!black}{\textbf{\ding{51}}}}
\newcommand{\bad}{\textcolor{red!70!black}{\textbf{\ding{55}}}}
\newcommand{\gpos}[1]{\textcolor{green!50!black}{#1}}
\newcommand{\rneg}[1]{\textcolor{red!70!black}{#1}}

\tikzset{
  masnode/.style={
    draw,
    rounded corners=2pt,
    fill=white,
    font=\tiny,
    align=center,
    inner sep=2pt,
    minimum height=3.8mm,
    text width=1.05cm
  },
  maswide/.style={masnode,text width=1.30cm},
  io/.style={masnode,fill=gray!12,text width=0.60cm},
  skillnode/.style={
    draw,
    rounded corners=2pt,
    fill=white,
    font=\scriptsize,
    align=center,
    inner sep=2pt,
    minimum height=4.4mm,
    text width=1.55cm
  },
  skillwide/.style={skillnode,text width=2.05cm},
  skillio/.style={skillnode,fill=gray!12,text width=0.80cm}
}

\newcommand{\diagEvoBCP}{%
\begin{tikzpicture}[baseline=(current bounding box.center), scale=0.44, >=Stealth, thick]
  \node[io] (in) at (0,0) {Input};
  \node[masnode] (meta) at (3.0,0) {Meta\\Planner};
  \node[masnode] (check) at (6.5,0) {Check\\Planner};
  \node[masnode] (e1) at (9.9,2.0) {Expert-1\\Retrieve};
  \node[masnode] (e2) at (9.9,0) {Expert-2\\Retrieve};
  \node[masnode] (e3) at (9.9,-2.0) {Expert-3\\Retrieve};
  \node[masnode] (pk) at (13.5,1.2) {PK / All\\Selection};
  \node[masnode] (ref) at (13.5,-1.2) {Refine\\Planner};
  \node[io] (out) at (16.2,0) {Output};
  \draw[->] (in.east)--(meta.west);
  \draw[->] (meta.east)--(check.west);
  \draw[->] (check.east)--(e1.west); \draw[->] (check.east)--(e2.west); \draw[->] (check.east)--(e3.west);
  \draw[->] (e1.east)--(pk.west); \draw[->] (e2.east)--(pk.west); \draw[->] (e3.east)--(pk.west);
  \draw[->] (pk.south)--(ref.north);
  \draw[->] (ref.east)--(out.west);
\end{tikzpicture}}

\newcommand{\diagEvoVita}{%
\begin{tikzpicture}[baseline=(current bounding box.center), scale=0.44, >=Stealth, thick]
  \node[io] (in) at (0,0) {Input};
  \node[masnode] (meta) at (3.0,0) {Meta\\Planner};
  \node[masnode] (check) at (6.5,0) {Check\\Planner};
  \node[masnode] (e1) at (9.9,2.2) {Expert-1\\Vita\\Tools Loop};
  \node[masnode] (e2) at (9.9,0) {Expert-2\\Vita\\Tools Loop};
  \node[masnode] (e3) at (9.9,-2.2) {Expert-3\\Vita\\Tools Loop};
  \node[masnode] (pk) at (13.5,1.2) {PK / All\\Selection};
  \node[masnode] (ref) at (13.5,-1.2) {Refine\\Planner};
  \node[io] (out) at (16.2,0) {Output};
  \draw[->] (in.east)--(meta.west);
  \draw[->] (meta.east)--(check.west);
  \draw[->] (check.east)--(e1.west); \draw[->] (check.east)--(e2.west); \draw[->] (check.east)--(e3.west);
  \draw[->] (e1.east)--(pk.west); \draw[->] (e2.east)--(pk.west); \draw[->] (e3.east)--(pk.west);
  \draw[->] (pk.south)--(ref.north);
  \draw[->] (ref.east)--(out.west);
\end{tikzpicture}}

\newcommand{\diagAOrchBCP}{%
\begin{tikzpicture}[baseline=(current bounding box.center), scale=0.44, >=Stealth, thick]
  \node[io] (in) at (0,0) {Input};
  \node[maswide] (main) at (3.3,0) {MainAgent\\Orchestrator};
  \node[masnode] (search) at (6.9,2.0) {BrowseComp\\Search};
  \node[masnode] (delegate) at (6.9,-1.8) {Delegate\\Task Tool};
  \node[masnode] (sub) at (10.4,-1.8) {SubAgent\\(TextOnly\\ / ReAct)};
  \node[masnode] (complete) at (13.8,-0.2) {Complete\\Action};
  \node[io] (out) at (16.7,-0.2) {Output};
  \draw[->] (in.east)--(main.west);
  \draw[->] (main.east)--(search.west);
  \draw[->] (main.east)--(delegate.west);
  \draw[->] (search.south)--(delegate.north);
  \draw[->] (delegate.east)--(sub.west);
  \draw[->] (sub.east)--(complete.west);
  \draw[->] (complete.east)--(out.west);
\end{tikzpicture}}

\newcommand{\diagAOrchVita}{%
\begin{tikzpicture}[baseline=(current bounding box.center), scale=0.44, >=Stealth, thick]
  \node[io] (in) at (0,0) {Input};
  \node[maswide] (main) at (3.2,0) {MainAgent\\Orchestrator};
  \node[masnode] (delegate) at (6.6,-1.6) {Delegate\\Task Tool};
  \node[masnode] (sub) at (10.3,-1.6) {SubAgent};
  \node[masnode] (vita) at (6.6,1.8) {Vita Env\\Tool Bridge};
  \node[masnode] (complete) at (13.6,-0.2) {Answer\\Complete};
  \node[io] (out) at (16.4,-0.2) {Output};
  \draw[->] (in.east)--(main.west);
  \draw[->] (main.east)--(vita.west);
  \draw[->] (vita.south)--(delegate.north);
  \draw[->] (delegate.east)--(sub.west);
  \draw[->] (sub.east)--(complete.west);
  \draw[->] (complete.east)--(out.west);
\end{tikzpicture}}

\newcommand{\diagAFlowBCP}{%
\begin{tikzpicture}[baseline=(current bounding box.center), scale=0.44, >=Stealth, thick]
  \node[io] (in) at (0,0) {Input};
  \node[masnode] (g1) at (3.0,2.2) {Search-1\\BCP};
  \node[masnode] (g2) at (3.0,0) {Search-2\\BCP};
  \node[masnode] (g3) at (3.0,-2.2) {Search-3\\BCP};
  \node[masnode] (ens) at (6.8,0) {SC\\Ensemble};
  \node[masnode] (ver) at (10.4,0) {Answer\\Generate\\Reasoning};
  \node[masnode] (fmt) at (14.0,0) {Custom\\Format};
  \node[io] (out) at (17.1,0) {Output};
  \draw[->] (in.east)--(g1.west); \draw[->] (in.east)--(g2.west); \draw[->] (in.east)--(g3.west);
  \draw[->] (g1.east)--(ens.west); \draw[->] (g2.east)--(ens.west); \draw[->] (g3.east)--(ens.west);
  \draw[->] (ens.east)--(ver.west);
  \draw[->] (ver.east)--(fmt.west);
  \draw[->] (fmt.east)--(out.west);
\end{tikzpicture}}

\newcommand{\diagAFlowVita}{%
\begin{tikzpicture}[baseline=(current bounding box.center), scale=0.44, >=Stealth, thick]
  \node[io] (in) at (0,0) {Input};
  \node[masnode] (gen) at (3.6,0) {Custom\\GENERATE};
  \node[masnode] (ver) at (8.2,0) {Custom\\VERIFY};
  \node[io] (out) at (12.0,0) {Output};
  \draw[->] (in.east)--(gen.west);
  \draw[->] (gen.east)--(ver.west);
  \draw[->] (ver.east)--(out.west);
\end{tikzpicture}}

\newcommand{\diagMASSVita}{%
\begin{tikzpicture}[baseline=(current bounding box.center), scale=0.44, >=Stealth, thick]
  \node[io] (in) at (0,0) {Input};
  \node[maswide] (decomp) at (3.2,0) {Decomposition\\Plan};
  \node[masnode] (first) at (6.8,0) {First\\Solution};
  \node[masnode] (ref) at (10.4,0) {Refined\\Solution};
  \node[masnode] (ens) at (14.0,0) {Ensemble\\Pick and Tool};
  \node[io] (out) at (16.8,0) {Output};
  \draw[->] (in.east)--(decomp.west);
  \draw[->] (decomp.east)--(first.west);
  \draw[->] (first.east)--(ref.west);
  \draw[->] (ref.east)--(ens.west);
  \draw[->] (ens.east)--(out.west);
\end{tikzpicture}}

\newcommand{\diagOrchBCP}{%
\begin{tikzpicture}[baseline=(current bounding box.center), scale=0.40, >=Stealth, thick]
  \node[io] (in) at (0,0) {Input};
  \node[masnode] (search) at (3.5,0) {Search};
  \node[masnode] (team) at (8,4.0) {TEAM};
  \node[masnode] (father) at (8,2.4) {FATHER\_\\GOALS};
  \node[masnode] (tour) at (8,0.8) {TOURNAMENT};
  \node[masnode] (league) at (8,-0.8) {PRO\_\\LEAGUE};
  \node[masnode] (ngoals) at (8,-2.4) {NATIONAL\_\\GOALS};
  \node[masnode] (nevent) at (8,-4.0) {NATIONAL\_\\EVENT};
  \node[masnode] (final) at (14.0,0) {FINAL};
  \node[io] (out) at (17.4,0) {Output};
  \draw[->] (in.east)--(search.west);
  \draw[->] (search.east)--(team.west);
  \draw[->] (search.east)--(father.west);
  \draw[->] (search.east)--(tour.west);
  \draw[->] (search.east)--(league.west);
  \draw[->] (search.east)--(ngoals.west);
  \draw[->] (search.east)--(nevent.west);
  \draw[->] (team.east)--(final.west);
  \draw[->] (father.east)--(final.west);
  \draw[->] (tour.east)--(final.west);
  \draw[->] (league.east)--(final.west);
  \draw[->] (ngoals.east)--(final.west);
  \draw[->] (nevent.east)--(final.west);
  \draw[->] (final.east)--(out.west);
\end{tikzpicture}}

\newcommand{\diagOrchVita}{%
\begin{tikzpicture}[baseline=(current bounding box.center), scale=0.40, >=Stealth, thick]
  \node[io] (in) at (0,0) {Input};
  \node[masnode] (c1) at (4.8,2.0) {CoT-1};
  \node[masnode] (c2) at (4.8,1.0) {CoT-2};
  \node[masnode] (c3) at (4.8,0) {CoT-3};
  \node[masnode] (c4) at (4.8,-1.0) {CoT-4};
  \node[masnode] (c5) at (4.8,-2.0) {CoT-5};
  \node[masnode] (vote) at (9.8,0) {Majority\\Vote};
  \node[io] (out) at (13.0,0) {Output};
  \draw[->] (in.east)--(c1.west); \draw[->] (in.east)--(c2.west);
  \draw[->] (in.east)--(c3.west); \draw[->] (in.east)--(c4.west);
  \draw[->] (in.east)--(c5.west);
  \draw[->] (c1.east)--(vote.west); \draw[->] (c2.east)--(vote.west);
  \draw[->] (c3.east)--(vote.west); \draw[->] (c4.east)--(vote.west);
  \draw[->] (c5.east)--(vote.west);
  \draw[->] (vote.east)--(out.west);
\end{tikzpicture}}
\newcommand{\diagInitBCP}{%
\begin{tikzpicture}[baseline=(current bounding box.center), scale=0.40, >=Stealth, thick]
  \node[skillio] (in) at (0,3) {Input};
  \node[skillnode,fill=blue!8] (h) at (4.5,3) {hint\_parser};
  \node[skillnode,fill=green!10] (t) at (10.0,3) {team\_career\_\\retriever};
  \node[skillnode,fill=green!10] (s) at (16.0,3) {statistics\_\\filter};
  \node[skillnode,fill=green!10] (a) at (16.0,0) {award\_title\_\\checker};
  \node[skillnode,fill=orange!12] (m) at (10.0,0) {media\_feature\_\\retriever};
  \node[skillnode,fill=red!8] (u) at (4.5,0) {unique\_event\_\\validator};
  \node[skillio] (out) at (0,0) {Output};
  \draw[->] (in.east)--(h.west);
  \draw[->] (h.east)--(t.west);
  \draw[->] (t.east)--(s.west);
  \draw[->] (s.south)--(a.north);
  \draw[->] (a.west)--(m.east);
  \draw[->] (m.west)--(u.east);
  \draw[->] (u.west)--(out.east);
\end{tikzpicture}}

\newcommand{\diagOptBCP}{%
\begin{tikzpicture}[baseline=(current bounding box.center), scale=0.40, >=Stealth, thick]
  \node[skillio] (in) at (0,3) {Input};
  \node[skillwide,fill=blue!8] (p) at (0,0) {parse\_and\_plan};
  \node[skillnode,fill=green!10] (r1) at (8.0,4.0) {retrieve\_team\_\\player\_pair};
  \node[skillnode,fill=green!10] (r2) at (8.0,2.0) {retrieve\_\\tournament\_stats};
  \node[skillnode,fill=green!10] (r3) at (8.0,0.0) {retrieve\_pro\_\\league\_title};
  \node[skillnode,fill=green!10] (r4) at (8.0,-2.0) {retrieve\_national\_\\and\_goals};
  \node[skillnode,fill=green!10] (r5) at (8.0,-4.0) {retrieve\_national\_\\appearance};
  \node[skillnode,fill=orange!12] (v) at (14,0) {link\_verification};
  \node[skillnode,fill=red!8] (m) at (14,-2) {merge\_and\_decide};
  \node[skillio] (out) at (14,-4) {Output};
  \draw[->] (in.south)--(p.north);
  \draw[->] (p.east)--(r1.west); \draw[->] (p.east)--(r2.west); \draw[->] (p.east)--(r3.west); \draw[->] (p.east)--(r4.west); \draw[->] (p.east)--(r5.west);
  \draw[->] (r1.east)--(v.west); \draw[->] (r2.east)--(v.west); \draw[->] (r3.east)--(v.west); \draw[->] (r4.east)--(v.west); \draw[->] (r5.east)--(v.west);
  \draw[->] (v.south)--(m.north);
  \draw[->] (m.south)--(out.north);
\end{tikzpicture}}

\newcommand{\diagInitVita}{%
\begin{tikzpicture}[baseline=(current bounding box.center), scale=0.40, >=Stealth, thick]
  \node[skillio] (in) at (0,2) {Input};
  \node[skillnode,fill=green!10] (o) at (5.0,2) {order\_meal};
  \node[skillnode,fill=green!10] (b) at (10.8,2) {find\_book\_cafe};
  \node[skillnode,fill=green!10] (v) at (16.8,2) {purchase\_voucher};
  \node[skillnode,fill=red!8] (t) at (16.8,0) {book\_train\_ticket};
  \node[skillio] (out) at (10.8,0) {Output};
  \draw[->] (in.east)--(o.west);
  \draw[->] (o.east)--(b.west);
  \draw[->] (b.east)--(v.west);
  \draw[->] (v.south)--(t.north);
  \draw[->] (t.west)--(out.east);
\end{tikzpicture}}

\newcommand{\diagOptVita}{%
\begin{tikzpicture}[baseline=(current bounding box.center), scale=0.40, >=Stealth, thick]
  \node[skillio] (in) at (0,3) {Input};
  \node[skillnode,fill=blue!8] (c) at (0,0) {constraint\_\\extraction};
  \node[skillnode,fill=green!10] (l1) at (5.0,3.6) {light\_food\_\\exploration};
  \node[skillnode,fill=green!10] (l2) at (9.9,3.6) {light\_food\_\\evaluation};
  \node[skillnode,fill=green!10] (l3) at (14.8,3.6) {light\_food\_\\ordering};
  \node[skillnode,fill=green!10] (b1) at (5.0,1.2) {bookbar\_\\exploration};
  \node[skillnode,fill=green!10] (b2) at (9.9,1.2) {bookbar\_\\evaluation};
  \node[skillnode,fill=green!10] (b3) at (14.8,1.2) {bookbar\_voucher\_\\purchase};
  \node[skillnode,fill=green!10] (t1) at (5.0,-1.2) {train\_ticket\_\\exploration};
  \node[skillnode,fill=green!10] (t2) at (9.9,-1.2) {train\_ticket\_\\evaluation};
  \node[skillnode,fill=green!10] (t3) at (14.8,-1.2) {train\_ticket\_\\purchase};
  \node[skillnode,fill=red!8] (f) at (15.8,-3.5) {final\_summary};
  \node[skillio] (out) at (10.8,-3.5) {Output};
  \draw[->] (in.south)--(c.north);
  \draw[->] (c.east)--(l1.west); \draw[->] (c.east)--(b1.west); \draw[->] (c.east)--(t1.west);
  \draw[->] (l1.east)--(l2.west); \draw[->] (l2.east)--(l3.west);
  \draw[->] (b1.east)--(b2.west); \draw[->] (b2.east)--(b3.west);
  \draw[->] (t1.east)--(t2.west); \draw[->] (t2.east)--(t3.west);
  \draw[->] (l3.east)--(f.east);
  \draw[->] (b3.east)--(f.east);
  \draw[->] (t3.east)--(f.east);
  \draw[->] (f.west)--(out.east);
\end{tikzpicture}}

\usepackage[T1]{fontenc}

\usepackage[utf8]{inputenc}

\usepackage{microtype}

\usepackage{inconsolata}

\usepackage{graphicx}

\newcommand{\emnlpcr}[1]{\textcolor{black}{#1}}
\title{Skill-MAS: Evolving Meta-Skill for Automatic Multi-Agent Systems}

\author{
\bf Hehai Lin\textsuperscript{$^{\spadesuit}$}, ~
Qi Yang\textsuperscript{$^{\diamondsuit}$}, ~
Chengwei Qin\textsuperscript{$^{\spadesuit}$} \thanks{Corresponding to \href{mailto:chengweiqin@hkust-gz.edu.cn}{chengweiqin@hkust-gz.edu.cn}}~
\\
$^{\diamondsuit}$Ant Group,
$^{\spadesuit}$The Hong Kong University of Science and Technology (Guangzhou)
}

\begin{document}
\maketitle

\begin{strip} 
\vspace{-9em} 
\begin{center}
\twemoji{brain} \href{https://linhh29.github.io/blog/Skill-MAS/index.html}{Project page} \hspace{0.5em} \textcolor{red}{\faYoutube} \href{https://skill-mas-demo.hehailin.life/}{Gallery \& Live Demo} \hspace{0.5em} \faGithub\ \href{https://github.com/linhh29/Skill-MAS}{Code}
\end{center}
\end{strip} 

\begin{abstract}
Large Language Model (LLM)-based automatic Multi-Agent Systems (MAS) generation has become a crucial frontier for tackling complex tasks. However, existing methods face a dilemma between \textbf{model capability} and \textbf{experience retention}. Inference-time MAS leverages frozen frontier LLMs but repeats identical searches without learning from past experience. Conversely, Training-time MAS internalizes experience via gradient updates but is constrained by the low capability ceiling of smaller models, and is hard to scale to large frontier LLMs. To bridge this gap, we propose Skill-MAS, a novel third path that decouples experience retention from parametric updates by conceptualizing the high-level orchestration capability as an evolvable Meta-Skill. Skill-MAS refines this architectural knowledge through a closed optimization loop: (1) \textbf{Multi-Trajectory Rollout} samples a behavioral distribution for each task under the current Meta-Skill; and (2) \textbf{Selective Reflection} adaptively selects priority tasks and applies hierarchical contrastive analysis to distill systemic experience into generalizable, strategy-level principles. Extensive experiments across four complex benchmarks and four distinct LLMs demonstrate that Skill-MAS not only achieves remarkable performance gains but also maintains a favorable cost-performance trade-off. Further analysis reveals that the evolved Meta-Skills are highly robust and exhibit strong transferability across unseen tasks and different LLMs.
\end{abstract}

\section{Introduction}
\label{sec:introduction}
Large Language Model (LLM)-based Multi-Agent Systems (MAS) have demonstrated remarkable efficacy in tackling complex tasks through collaboration~\cite{xu2025toward, lin2025interactive, wu2025furina,huang2026ama}. However, manually designing specific agent roles, communication topologies, and workflows for highly heterogeneous tasks is labor-intensive and challenging to scale~\cite{chen2025enhancing,lin2026unified}. Consequently, automatic-MAS has emerged as a pivotal direction, aiming to automate the generation and optimization of multi-agent architectures~\cite{ye2025mas, tran2025multi, ke2025survey, jwalapuram2026illusion}.

\begin{figure*}[ht]
    \centering
    \includegraphics[width=\linewidth]{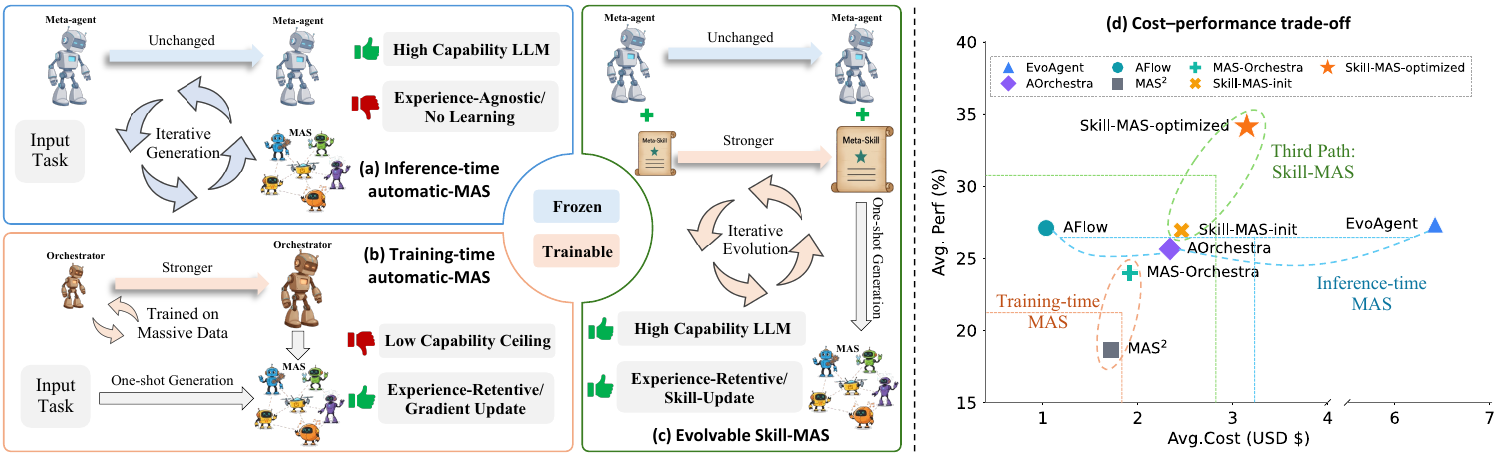} 
    \caption{Overview of MAS paradigms. (a)-(b) Comparison of existing Inference-time and Training-time MAS, illustrating the dilemma between model capability and experience retention. (c) Evolvable Skill-MAS bridges this gap by iteratively evolving the Meta-Skill, successfully coupling high-capability LLMs with experience retentiveness. (d) Cost-performance trade-off analysis, highlighting Skill-MAS as a better ``Third Path'' for automatic-MAS.}

    \label{fig:background}

\end{figure*}

Based on how orchestration knowledge is acquired and retained, existing automatic-MAS predominantly split into two distinct tracks\footnote{Meta-agent refers to inference-time MAS generator, and Orchestra denotes training-time MAS generator.}, each presenting a trade-off between \textbf{model capability} and \textbf{experience retention}. 
The first track is inference-time orchestration~\cite{ke2025mas, ruan2026aorchestra}, which couples a frozen frontier Meta-agent with iterative search algorithms to optimize MAS. While this approach benefits from the strong reasoning capability of state-of-the-art LLMs, its optimization process is inherently \textit{experience-agnostic}. The Meta-agent operates without a cumulative memory mechanism. It repeats identical search and generation procedures across different runs, failing to retain or transfer valuable diagnosis experiences learned from prior failures~\cite{yang2026agentnet}. 
The second track is training-time orchestration~\cite{su2025toolorchestra, ke2026mas}, which parameterizes the orchestration capability by fine-tuning smaller LLMs (typically 7B parameters) on curated orchestration datasets. Although this track enables the model to internalize experiences through gradient updates, it is constrained by the low capability ceiling of smaller LLMs. Furthermore, it requires massive high-quality training data, which cannot easily scale to or leverage proprietary, ultra-large frontier models ($>$100B) where the strongest reasoning capacities reside~\cite{rank2026posttrainbench}. 
Consequently, this dilemma raises a crucial question: Is there a third path that decouples experience retention from parametric updates, thereby enabling a frozen frontier Meta-agent to progressively learn and refine its orchestration expertise across tasks?

The emerging ``Skill'' paradigm within the community offers a potential solution to this question~\cite{xiong2026ace, si2026context, vishe2026skill}. By abstracting capabilities into structured, natural language documentation (e.g., SKILL.md), agents can achieve progressive enhancement of such skills through iterative evolution~\cite{zhang2026evoskills, li2026arise}. However, existing efforts in skill discovery and evolution are heavily concentrated on single-agent scenarios. For example, MemSkill iteratively mines and refines memory management skills for memory agents~\cite{zhang2026memskill}. Even when extended to MAS, skill analysis remains strictly confined to the sub-agent level, focusing solely on how individual task-executing agents within a complex system can be abstracted into corresponding skills~\cite{li2026single, alzubi2026evoskill}. 
However, we observe that the Meta-agent's high-level orchestration behavior can similarly be modeled as an evolvable skill (\S\ref{subsec:skill_formulation}). By adopting this perspective, we can successfully endow the Meta-agent with the capacity for iterative learning.

To this end, we propose \textbf{Skill-MAS} that conceptualizes the Meta-agent's orchestration capability as an evolvable Meta-Skill, encoding strategy-level principles spanning task decomposition, agent engineering, and workflow orchestration. Then Skill-MAS refines this Meta-Skill through a closed optimization loop comprising two stages. First, \textit{Multi-Trajectory Rollout} executes multiple independent rollouts per task under the current skill, converting single-trial outcomes into distributional statistics that separate genuine capability from execution stochasticity. Second, \textit{Selective Reflection} prioritizes the most volatile and difficult tasks via a joint uncertainty--difficulty score with adaptive elbow truncation. Then it applies hierarchical contrastive analysis (within-task then cross-task) to diagnose systemic failure modes, and feeds the resulting evidence to optimize the Meta-Skill while preserving its three-module scaffold. After rounds, the best-performing skill is selected for final evaluation.

To validate the effectiveness of Skill-MAS, we conduct evaluations across four challenging benchmarks, spanning deep research, expert-level mathematics, multi-hop question answering, and real-world interactive scenarios. Utilizing four LLMs as the Meta-agent, we compare Skill-MAS against state-of-the-art Inference-time and Training-time automatic-MAS. Experimental results demonstrate that Skill-MAS delivers remarkable performance while maintaining a better cost-performance trade-off. Our further analysis reveals that the evolved Meta-Skills encode generalizable orchestration strategies and exhibit robust transferability across both unseen tasks and different backbone LLMs. These findings collaboratively highlight the importance of our novel paradigm: augmenting a static Meta-agent with continuously evolving Meta-Skill.
Our main contributions are summarized as follows:

\begin{itemize}
    \item We pioneer a novel automatic-MAS paradigm that conceptualizes high-level orchestration behavior as an evolvable Meta-Skill, enabling frozen frontier LLMs to refine architectural knowledge without costly parametric updates.
    \item We propose Skill-MAS to iteratively evolve the Meta-Skill via Multi-Trajectory Rollout and Selective Reflection, distilling experience into generalizable orchestration principles.
    \item Extensive experiments demonstrate that our Skill-MAS achieves a better cost-performance trade-off with highly transferable Meta-Skills.
\end{itemize}

\section{Related Work}
\label{sec:relatedwork}


\subsection{Automatic-MAS}
\textbf{Inference-time} approaches leverage the strong reasoning capabilities of frozen frontier LLMs to serve as the Meta-agent~\cite{wang2025megaagent, zhang2025metaagent}. By coupling these models with sophisticated prompts and search algorithms, these methods aim to discover effective MAS without altering the underlying model weights~\cite{ferrag2025llm}. For example,
ADAS represents the MAS as executable code and iteratively proposes new architectures through meta-agent search on validation~\cite{hu2024automated}, while 
AFlow employs search algorithms like Monte Carlo Tree Search to navigate the expansive space of agentic workflows~\cite{zhang2024aflow}. Concurrently, another subset of works eliminates the need for a validation set. For instance, EvoAgent extends individual expert agents into multi-agent collaborative networks through evolutionary principles~\cite{yuan2025evoagent}. AOrchestra dynamically populates node attributes based on hierarchical task decomposition~\cite{ruan2026aorchestra}. MAS-Zero introduces self-reflective feedback loops to refine the MAS topology~\cite{ke2025mas}.
Within these frameworks, the Meta-agent executes identical search routines in every iteration and is structurally incapable of achieving experiential learning across historical orchestration trajectories~\cite{li2024survey, he2025self}.

\textbf{Training-time} approaches teach the orchestrator (a small LLM) to generate multi-agent configurations in a single pass by learning from historical trajectories~\cite{ye2025mas, su2025toolorchestra}. For example, ScoreFlow utilizes a variant of direct preference optimization to inject quantitative feedback directly into the training process~\cite{wang2025scoreflow}. MAS$^2$ trains models to master self-generative and self-rectifying workflows~\cite{wang2025mas}, while MAS-Orchestra models the MAS construction as a sequential function-calling task optimized via GRPO~\cite{ke2026mas}.
Although this approach provides a learning pathway for mastering orchestration capabilities, the parametric optimization requires an extensive collection of complex orchestration trajectories~\cite{dang2026multi}. More crucially, this paradigm is prohibitively difficult to scale up to those powerful frontier LLMs ($>$100B) because of the astronomical computational costs associated with fine-tuning or reinforcement learning~\cite{rank2026posttrainbench}.

Unlike these two categories, Skill-MAS pioneers a third path: treating the Meta-agent's orchestration as an evolvable Meta-Skill, progressively empowering the Meta-agent to generate superior MAS.

\subsection{Skill Evolution and Analysis}
The emerging ``Skill'' paradigm offers a powerful mechanism for agent self-improvement by abstracting operational capabilities into structured, natural language documentation~\cite{yang2026autoskill}. This allows agents to distill experience into reusable principles through evolution~\cite{wu2026co}.
Current literature primarily focuses on evolving execution-level skills within single-agent scenarios. Frameworks such as MemSkill~\cite{zhang2026memskill} and Trace2Skill~\cite{ni2026trace2skill} emphasize the extraction of task-specific routines by analyzing historical interaction traces, ranging from memory management to multi-hop reasoning. Other methodologies, including Skill0~\cite{lu2026skill0}, D2Skill~\cite{tu2026dynamic}, and SKILLRL~\cite{xia2026skillrl}, integrate skill discovery with policy optimization, allowing agents to internalize heuristics from environmental feedback. When the skill paradigm is extended to multi-agent systems, research still focuses on the sub-agent level~\cite{pan2026skillmas}. For example, CoEvoSkills~\cite{zhang2026evoskills} and EvoSkill~\cite{alzubi2026evoskill} iteratively refine the executable roles of task-executing agents, while recent studies on role-based skill libraries investigate how to organize these sub-agent skills via hierarchical routing~\cite{li2026single}. 

Different from previous works that focus on the evolution of execution-level skills, we conceptualize the Meta-agent’s high-level orchestration behavior as a Meta-Skill, an evolvable artifact that captures meta-level architectural know-how. 



\section{Methodology}
\label{sec:methodology}

\begin{figure*}[t]
    \centering
    \includegraphics[width=\linewidth]{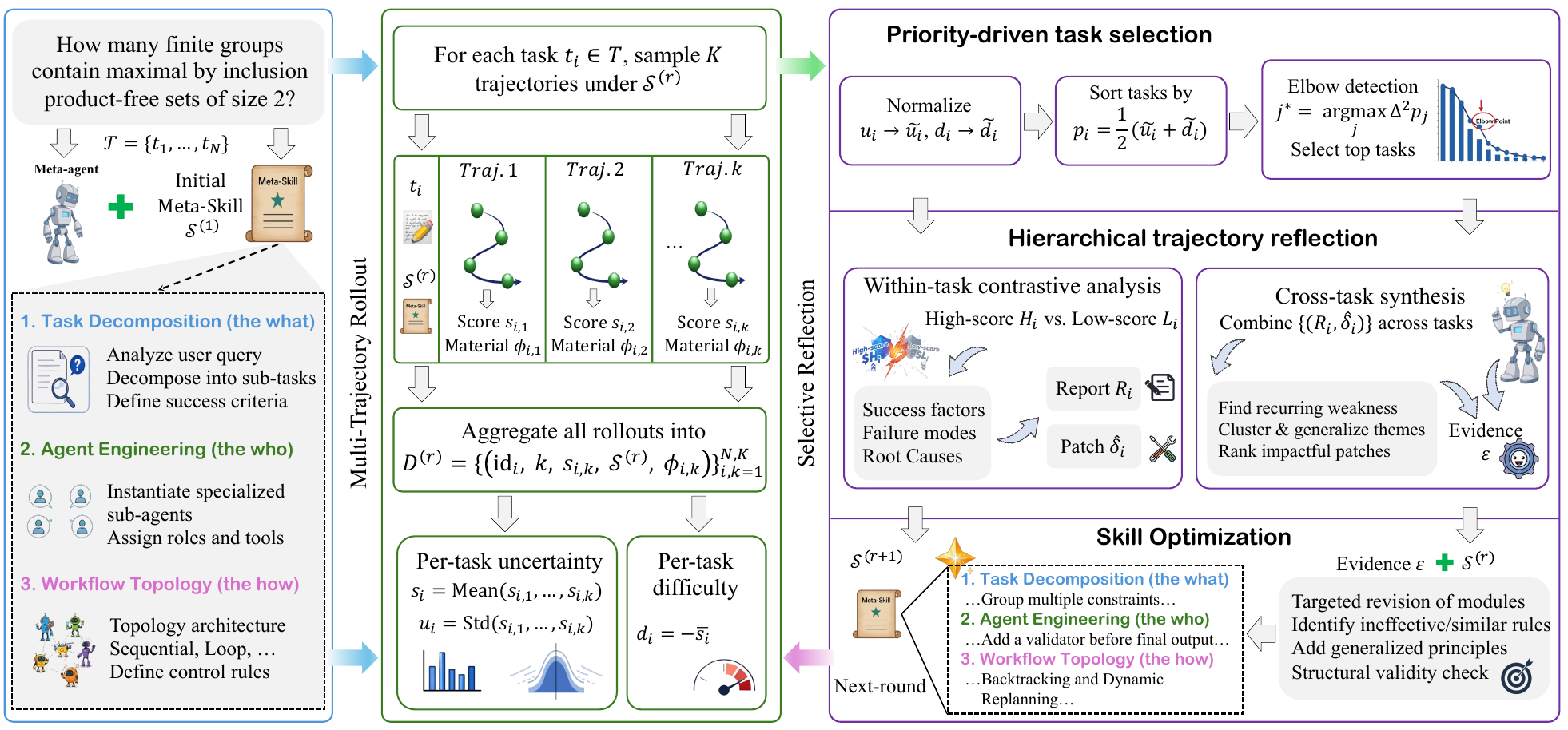} 
    \caption{The evolutionary loop of Skill-MAS. The Meta-Skill $\mathcal{S}^{(r)}$ (task decomposition, agent engineering, and workflow topology) guides Multi-Trajectory Rollout to compute distributional statistics. These metrics feed into Selective Reflection to prioritize tasks and extract evidence $\mathcal{E}$ via hierarchical trajectory reflection. Skill Optimization then leverages $\mathcal{E}$ to refine the Meta-Skill into $\mathcal{S}^{(r+1)}$, driving iterative improvement in MAS generation.}
    \label{fig:method}
\end{figure*}

We formulate the Meta-agent's orchestration as a Meta-Skill. Let $\mathcal{T} = \{t_1,\ldots,t_N\}$ denote the validation set and $\mathcal{S}^{(r)}$ represents the Meta-Skill active in round $r \in \{1,\ldots,R\}$. Each round proceeds in two stages: \textit{Multi-Trajectory Rollout} (\S\ref{subsec:rollout}) samples a behavioral distribution over $\mathcal{T}$ under $\mathcal{S}^{(r)}$, and \textit{Selective Reflection} (\S\ref{subsec:reflect}) diagnoses high-priority failure modes
and rewrites the skill into $\mathcal{S}^{(r+1)}$. After $R$ rounds, we select the validation-optimal skill as $\mathcal{S}^{*}$ for final test-set evaluation.


\subsection{Meta-Skill Formulation}
\label{subsec:skill_formulation}

We model the Meta-agent's orchestration behavior as a structured skill $\mathcal{S}$. Rather than encoding task-specific solutions, $\mathcal{S}$ captures reusable, \textit{strategy-level} principles that govern how the Meta-agent constructs a MAS for an arbitrary query. 

The skill $\mathcal{S}$ is organized into three modules that collectively span the full orchestration pipeline.
(1) \textbf{Task Decomposition} (the \textit{what}) prescribes how the Meta-agent analyzes a user query: it identifies the macro objective and scope, decomposes the request into discrete, logically cohesive sub-tasks, and specifies evaluable success criteria for each sub-task.
(2) \textbf{Agent Engineering} (the \textit{who}) governs the instantiation of specialized sub-agents: each is assigned a distinct role profile and provided with the specific contextual inputs it requires to operate.
(3) \textbf{Workflow Orchestration} (the \textit{how}) instructs the Meta-agent to select an appropriate architectural topology (e.g., sequential, hierarchical, or loop) to define precise input-output mappings across agents and to emit an executable MAS.

This tripartite formulation serves a dual purpose: at inference time, it provides the Meta-agent with a principled procedure for MAS generation; during optimization, it enables each round's diagnostic phase to precisely attribute failures to a specific module, ensuring that skill updates remain localized and interpretable. We use LLM to summarize the initial Meta-Skill $\mathcal{S}^{(1)}$ (Appendix~\ref{appendix:case study}) from~\citet{anthropic} and subsequently apply an iterative optimization procedure to enhance its efficacy.

\subsection{Multi-Trajectory Rollout}
\label{subsec:rollout}

In each round $r$, we sample $K$ trajectories independently for each task $t_i \in \mathcal{T}$ under the Meta-Skill $\mathcal{S}^{(r)}$. Each rollout traverses the complete three-module pipeline and produces a terminal outcome together with a full execution trace, which is recorded as a standardized trajectory:
\begin{align}
\tau_{i,k} = \bigl(\mathrm{id}_i,\; k,\; s_{i,k},\; \mathcal{S}^{(r)},\; \Phi_{i,k}\bigr)
\end{align}
where $\mathrm{id}_i$ identifies the task, $k$ indexes the trajectory, $s_{i,k} \in [0,1]$ is the normalized score, $\mathcal{S}^{(r)}$ is the Meta-Skill governing this round, and $\Phi_{i,k}$ stores the MAS's architecture and intermediate results. The round-$r$ corpus is $\mathcal{D}^{(r)} = \{\tau_{i,k}\}_{i=1,k=1}^{N,K}$.

From $\mathcal{D}^{(r)}$, two per-task statistics are derived to characterize the behavioral distribution. The \textit{uncertainty} of task $t_i$ is measured by the standard deviation of its $K$ trajectory scores:
\begin{align}
u_i = \sqrt{\frac{1}{K}\sum_{k=1}^{K}(s_{i,k}-\bar{s}_i)^2}, \quad \bar{s}_i = \frac{1}{K}\sum_{k=1}^{K} s_{i,k}
\end{align}
which quantifies how inconsistently the current skill orchestrates the same task across runs, and a large $u_i$ signals ambiguous or underspecified rules. 

The \textit{difficulty} of task $t_i$ is defined as the negated mean score:
$
d_i = -\bar{s}_i
$,
so that harder tasks (lower mean performance) receive larger values (the raw negative values are subsequently mapped to $[0,1]$ via min--max normalization in \S\ref{subsec:reflect}). Together, these two statistics convert each task from a single pass/fail outcome into a distributional characterization, enabling the subsequent reflection phase to distinguish systematic skill deficiencies from transient execution noise.

\subsection{Selective Reflection}
\label{subsec:reflect}

Given the rollout corpus $\mathcal{D}^{(r)}$ and its per-task statistics, the reflection stage transforms distributional evidence into targeted skill revisions. Rather than uniformly analyzing all $N$ tasks, which would dilute the diagnostic signal, we first perform \textit{priority-driven task selection} to isolate the most informative subset, then apply \textit{hierarchical trajectory reflection} to diagnose failure modes and produce actionable evidence for \textit{skill optimization}.

\subsubsection{Priority-Driven Task Selection} 
To concentrate the optimization budget, we fuse the two rollout statistics into a single priority score per task. Both $u_i$ and $d_i$ are first min--max normalized across all tasks within the current round:
\begin{align}
\tilde{v}_i = \frac{v_i - \min_j\, v_j}{\max_j\, v_j - \min_j\, v_j}, \quad v \in \{u,\, d\}
\end{align}
and then blended into a unified priority:
$
p_i = \frac{1}{2}(\tilde{u}_i + \tilde{d}_i)
$,
which jointly favors tasks that are both volatile and systematically difficult. Sorting by $p_i$ in descending order yields a priority curve $p_{(1)} \geq \cdots \geq p_{(N)}$. Since the priority values form a discrete sequence, we apply finite differences as the discrete analogue of derivatives to detect the curve's natural elbow. Let $\delta_j = p_{(j)} - p_{(j+1)}$ denote the first-order differences (analogous to the first derivative); the elbow index is then identified at the position of maximum absolute second-order difference (analogous to the second derivative):
\begin{align}
j^* = {\arg\max}_{j \in \{1,\ldots,N-2\}} \bigl|\delta_j - \delta_{j+1}\bigr| + 1
\end{align}
Then we apply hierarchical reflection on the selected trajectory set $\mathcal{T}_{\mathrm{sel}}= \{t_{(1)},\ldots,t_{(j^*)}\}$.

\subsubsection{Hierarchical Trajectory Reflection} 
The reflection operates in two phases to progressively lift task-specific observations into system-level diagnostic evidence: first analyzing trajectories \textit{within} each selected task, then synthesizing findings \textit{across} tasks.

\textbf{Phase~1: Within-task contrastive analysis.} For each $t_i \in \mathcal{T}_{\mathrm{sel}}$, the $K$ trajectories are partitioned into a high-scoring set $\mathcal{H}_i = \{\tau_{i,k} \mid s_{i,k} \geq \mathrm{median}(\{s_{i,k}\}_k)\}$ and a low-scoring set $\mathcal{L}_i$. An LLM-based reflector examines the execution snapshots $\Phi_{i,k}$ across both groups to perform a structured contrastive diagnosis: it identifies the specific divergence points where $\mathcal{H}_i$ and $\mathcal{L}_i$ begin to exhibit different orchestration decisions, extracts the success factors that characterize high-scoring runs, catalogues the recurring failure modes in low-scoring runs, and attributes corresponding root causes. The reflector consolidates these findings into a per-task summarization report $\mathcal{R}_i$ together with a candidate patch $\hat{\delta}_i$ that proposes a targeted, actionable modification to the implicated skill module.
\textbf{Phase~2: Cross-task synthesis.} To avoid overfitting, the per-task reports $\{\mathcal{R}_i\}_{t_i \in \mathcal{T}_{\mathrm{sel}}}$ are then jointly analyzed to extract patterns that transcend individual tasks. This phase identifies systemic weaknesses, failure modes recurring across multiple tasks, as well as systemic strengths representing robust orchestration strategies that should be preserved. It then formulates the structured evidence package $\mathcal{E}$ as a prioritized repair list by aggregating the candidate patches $\{\hat{\delta}_i\}$ and ranking them by expected impact and implementation feasibility.

\subsubsection{Skill Optimization}
The evidence package $\mathcal{E}$ and $\mathcal{S}^{(r)}$ drive the skill optimizer to produce $\mathcal{S}^{(r+1)}$. The optimizer first reviews the current skill against $\mathcal{E}$ to identify existing guidance that the evidence suggests is ineffective or counterproductive, and removes or rewrites such rules before introducing new ones. It then performs strategic optimization while strictly preserving the three-module scaffold. Modifications are targeted at the specific modules implicated by $\mathcal{E}$, whether task decomposition heuristics, agent engineering specifications, or workflow orchestration rules. Crucially, each modification must be grounded in the reflection evidence and abstracted into a generalizable orchestration principle rather than a task-specific fix to maintain generalization. 

The revised skill undergoes a structural validity check before acceptance, and once accepted, the $\mathcal{S}^{(r+1)}$ replaces the current skill as the active Meta-Skill for round $r{+}1$'s rollout. After all $R$ rounds are complete, the skill achieving the highest validation performance is selected as $\mathcal{S}^*$ and evaluated on the held-out test set.

\section{Experimental Setup}
\label{sec:experiments}

\begin{table*}[ht]
    \centering
    
    \resizebox{\textwidth}{!}{%

    \begin{tabular}{l | ccccc c | ccccc c}
    \toprule
    \multirow{2.5}{*}{\textbf{Method}} & \multicolumn{6}{c|}{\textbf{Gemini-3.1-Flash}} & \multicolumn{6}{c}{\textbf{GPT-5.4-Nano}} \\
    \cmidrule(lr){2-7} \cmidrule(lr){8-13}
     & DRB & HLE & BCP & VITA & Avg.Perf $\uparrow$ & Avg.Cost $\downarrow$ & DRB & HLE & BCP & VITA & Avg.Perf $\uparrow$ & Avg.Cost $\downarrow$ \\
    \midrule
    
    EvoAgent        & 40.76 & 8.93 & 16.07 & 10.71 & 19.12 & 8.20 & \textbf{52.91} & 14.29 & 23.81 & 8.33 & 24.83 & 6.26 \\
    AOrchestra   & 39.49 & 7.14 & 14.88 & 14.29 & 18.95 & 2.31 & 44.68 & 8.93 & 22.62 & 4.76 & 20.25 & 2.45 \\
    AFlow    & 42.31 & 5.95 & 17.86 & 19.05 & 21.29 & 0.44 & 44.29 & 11.31 & 18.45 & 2.38 & 19.11 & 0.69 \\
    MAS$^2$    & 38.03 & 2.33 & 11.39 & 8.33 & 15.02 & 0.91 & 43.20 & 10.42 & 10.71 & 2.38 & 16.68 & 1.27 \\
    MAS-Orchestra   & 40.40 & 17.26 & 8.90 & 9.52 & 19.02 & 2.48 & 43.83 & 14.88 & 7.74 & 9.52 & 18.99 & 1.84 \\
    \midrule
    
    Skill-MAS-init   & 36.10 & 14.88 & 19.05 & 16.67 & 21.68 & 2.63 & 40.45 & 12.50 & 19.64 & 5.95 & 19.64 & 4.24 \\

    Skill-MAS-optimized      & \textbf{44.71} & \textbf{21.43} & \textbf{23.21} & \textbf{28.60} & \textbf{29.49} & 2.82 & 48.90 & \textbf{18.45} & \textbf{27.38} & \textbf{15.48} & \textbf{27.55} & 5.22 \\
    
    \midrule


    \multirow{2.5}{*}{\textbf{Method}} & \multicolumn{6}{c|}{\textbf{Qwen3.5-Plus}} & \multicolumn{6}{c}{\textbf{DeepSeek-V4-Flash}} \\
    \cmidrule(lr){2-7} \cmidrule(lr){8-13}
     & DRB & HLE & BCP & VITA & Avg.Perf $\uparrow$ & Avg.Cost $\downarrow$ & DRB & HLE & BCP & VITA & Avg.Perf $\uparrow$ & Avg.Cost $\downarrow$ \\
    \midrule

    EvoAgent        & 47.02 & 27.38 & 15.48 & 34.52 & 31.10 & 7.91 & 49.10 & 9.52 & 16.67 & 61.90 & 34.30 & 3.33 \\
    AOrchestra   & 44.05 & 12.50 & 19.05  & 40.48 & 29.02 & 1.91 & 41.03 & 19.64 & 18.45 & 58.33 & 34.36 & 2.69 \\
    AFlow    & 46.17 & 25.00 & 12.50 & 45.24 & 32.23 & 1.56 & 49.93 & 21.43 & 15.48 & 55.95 & 35.70 & 1.45 \\
    MAS$^2$    & 40.87 & 16.56 & 11.52 & 22.62 & 22.89 & 4.04 & 35.31 & 6.67 & 4.92 & 33.33 & 20.06 & 0.63 \\
    MAS-Orchestra   & 40.43 & 21.43 & 10.12 & 35.71 & 26.92 & 1.35 & 40.63 & 20.24 & 15.72 & 47.62 & 31.05 & 1.98 \\
    \midrule
    
    Skill-MAS-init   & 48.88 & 24.40 & 20.24 & 36.90 & 32.61 & 1.83 & 44.99 & 20.83 & 15.48 & 53.57 & 33.72 & 1.14 \\
    Skill-MAS-optimized      & \textbf{51.28} & \textbf{29.76} & \textbf{23.81} & \textbf{48.80} & \textbf{38.41} & 2.43 & \textbf{51.69} & \textbf{26.79} & \textbf{22.62} & \textbf{63.10} & \textbf{41.05} & 2.12 \\
    
    \bottomrule
    \end{tabular}
    }
    
    \caption{Quantification comparison of Skill-MAS and baselines. Skill-MAS-init/optimized correspond to Skill-MAS with $\mathcal{S}^{(1)}$/$\mathcal{S}^{*}$.
    \textbf{DRB}: DeepResearchBench, \textbf{HLE}: Humanity's Last Exam-Math, \textbf{BCP}: BrowseComp-Plus, \textbf{VITA}: VitaBench. \textbf{Avg.} reports average performance and inference cost. \textbf{Bold} denotes the best result.}
    \label{tab:main_results}
\end{table*}

\paragraph{Benchmarks and Evaluation Metrics.}
We select four complex benchmarks spanning different scenarios.
(1) \textbf{DeepResearchBench}~\cite{du2025deepresearch} for deep research tasks. Performance is measured by comprehensiveness, insight, instruction-following, and readability. 
(2) \textbf{Humanity's Last Exam-Math}~\cite{phan2025humanity} for complex expert-level mathematical reasoning. Performance is measured by the accuracy.
(3) \textbf{BrowseComp-Plus}~\cite{chen2025browsecomp} for complex multi-hop dynamic question answering. Performance is measured by the accuracy. 
(4) \textbf{VitaBench}~\cite{he2025vitabench} simulates real-world daily scenarios requiring multi-tool calling. Responses are scored by a rubric-based evaluator to measure the success rate. 
Dataset statistics are provided in Appendix~\ref{appendix:statistics of benchmarks}.
All metrics are normalized to $[0, 100\%]$. We report the average performance and the average \textbf{inference cost} on the held-out test set (in USD \$). The difference between inference cost and training/evolution cost is detailed in Appendix~\ref{appendix:cost analysis}.
The LLM-as-a-Judge prompts are cataloged in Appendix~\ref{appendix:prompt details}.

\paragraph{Baselines.}
To ensure a rigorous and comprehensive evaluation, we compare our approach against five state-of-the-art automatic-MAS, categorized into two paradigms.
(1) \textbf{Inference-time MAS}: EvoAgent~\cite{yuan2025evoagent}, AOrchestra~\cite{ruan2026aorchestra}, and AFlow~\cite{zhang2024aflow}. 
(2) \textbf{Training-time MAS}: MAS$^2$~\cite{wang2025mas} and MAS-Orchestra~\cite{ke2026mas}. 
Detailed descriptions can be found in Appendix~\ref{appendix:automatic mas baselines}.
We compare Skill-MAS against these baselines with initial skill $\mathcal{S}^{(1)}$ and optimized skill $\mathcal{S}^{*}$.



\paragraph{Test Models.}
To ensure a fair and consistent evaluation, we utilize the \emph{same} LLM for all components within each automatic-MAS. Our evaluation spans four different models, comprising two proprietary models: Gemini-3.1-Flash~\cite{team2023gemini} and GPT-5.4-Nano~\cite{singh2025openai}, alongside two highly capable open-source models: Qwen3.5-Plus~\cite{qwen3.5} and DeepSeek-V4-Flash~\cite{deepseekai2026deepseekv4}. Detailed hyperparameters and model configurations are documented in Appendix~\ref{appendix: implementation details}, while the specific prompt crafted for Skill-MAS can be found in Appendix~\ref{appendix:prompt details}.

\section{Results and Analysis}
\label{sec:results}

\subsection{Main Results}
Table~\ref{tab:main_results} presents the quantitative comparison of our Skill-MAS and baselines across four distinct benchmarks using four different LLMs as Meta-agent. From a performance perspective, we observe that even without the iterative evolution, Skill-MAS with initial Meta-Skill (Skill-MAS-init) demonstrates competitive performance. In several scenarios, it achieves results comparable to or even slightly better than the best baselines. For example, with Gemini-3.1-Flash as Meta-agent, Skill-MAS-init slightly surpasses the best baseline AFlow (21.68 vs. 21.29). When it comes to Skill-MAS with optimized Meta-Skill (Skill-MAS-optimized), the only exception occurs on DeepResearchBench with GPT-5.4-Nano, where EvoAgent maintains a higher advantage (48.90 vs. 52.91). In all other cases, Skill-MAS-optimized outperforms all baseline methods by a large margin and achieves the highest average performance across all LLMs. The performance of these two variants collaboratively demonstrates the effectiveness of Skill-MAS.

The cost-performance trade-off in Figure~\ref{fig:background}~(d) reveals distinct distribution patterns among the three automatic-MAS categories on the test set. On average, Training-time MAS achieves the lowest cost but suffers from the worst performance. Despite using extensive training data to train a lightweight orchestrator, they tend to generate simplistic MAS at test time, lacking the generalization needed to adapt to problems with varying difficulty from different domains. Conversely, Inference-time MAS achieves a performance enhancement but incurs the highest inference cost. This highlights that while iterative MAS refinement works, re-executing it per sample is prohibitively expensive. In contrast, Skill-MAS achieves a better cost-performance trade-off. By generating the MAS in one shot rather than re-optimizing per sample iteratively, it attains the best performance at a moderate cost. This further underscores the importance of our novel paradigm: augmenting a static Meta-agent with continuously evolving Meta-Skill.

\begin{table}[t]
\centering
\resizebox{0.48\textwidth}{!}{%
\begin{tabular}{llllcc}
\toprule

\multicolumn{6}{c}{\textbf{No Transfer}} \\
\midrule
\multicolumn{2}{c}{\textbf{Skill Source}} & \multicolumn{2}{c}{\textbf{Test Setting}} & \multirow{2.5}{*}{\textbf{Score}} & \multirow{2.5}{*}{\textbf{$\Delta$}} \\
\cmidrule(lr){1-2}\cmidrule(lr){3-4}
\textbf{LLM} & \textbf{Task} & \textbf{LLM} & \textbf{Task} &  &  \\
\midrule
GPT-5.4-Nano      & BCP  & GPT-5.4-Nano      & BCP  & 27.38 & $\uparrow$7.74 \\
DeepSeek-V4-Flash & BCP  & DeepSeek-V4-Flash & BCP  & 22.62 & $\uparrow$7.14 \\
GPT-5.4-Nano      & VITA & GPT-5.4-Nano      & VITA & 15.48 & $\uparrow$9.53 \\
DeepSeek-V4-Flash & VITA & DeepSeek-V4-Flash & VITA & 63.10 & $\uparrow$9.53 \\
\midrule

\multicolumn{6}{c}{\textbf{Panel A: Cross-LLM transfer (same task)}} \\
\midrule
GPT-5.4-Nano      & BCP  & DeepSeek-V4-Flash & BCP  & 18.45 & $\uparrow$2.97 \\
DeepSeek-V4-Flash & BCP  & GPT-5.4-Nano      & BCP  & 24.40 & $\uparrow$4.76 \\
GPT-5.4-Nano      & VITA & DeepSeek-V4-Flash & VITA & 63.10 & $\uparrow$9.53 \\
DeepSeek-V4-Flash & VITA & GPT-5.4-Nano      & VITA & 14.29 & $\uparrow$8.34\\
\midrule

\multicolumn{6}{c}{\textbf{Panel B: Cross-Task transfer (same LLM)}} \\
\midrule
GPT-5.4-Nano      & BCP  & GPT-5.4-Nano      & VITA & 13.10 & $\uparrow$7.15 \\
GPT-5.4-Nano      & VITA & GPT-5.4-Nano      & BCP  & 23.21 & $\uparrow$3.57 \\
DeepSeek-V4-Flash & BCP  & DeepSeek-V4-Flash & VITA & 59.52 & $\uparrow$5.95 \\
DeepSeek-V4-Flash & VITA & DeepSeek-V4-Flash & BCP  & 20.83 & $\uparrow$5.35 \\
\midrule

\multicolumn{6}{c}{\textbf{Panel C: Cross-LLM + Cross-Task transfer}} \\
\midrule
GPT-5.4-Nano      & BCP  & DeepSeek-V4-Flash & VITA & 55.95 & $\uparrow$2.38 \\
GPT-5.4-Nano      & VITA & DeepSeek-V4-Flash & BCP  & 16.67 & $\uparrow$1.19 \\
DeepSeek-V4-Flash & BCP  & GPT-5.4-Nano      & VITA & 9.52 &  $\uparrow$3.57\\
DeepSeek-V4-Flash & VITA & GPT-5.4-Nano      & BCP  & 22.62 & $\uparrow$2.98 \\
\bottomrule
\end{tabular}
}
\caption{Transfer performance across LLMs and tasks. \textbf{Skill Source} denotes the (\textit{LLM, Task}) where the Meta-skill is evolved, and \textbf{Test Setting} denotes the (\textit{LLM, Task}) where that Meta-skill is evaluated. $\Delta$ is measured against Skill-MAS-init.}
\label{tab:transfer_three_panels}
\end{table}

\subsection{Further Analysis}

\subsubsection{Transferability of Meta-Skill}

\begin{figure}[t]
    \centering
    \includegraphics[width=\linewidth]{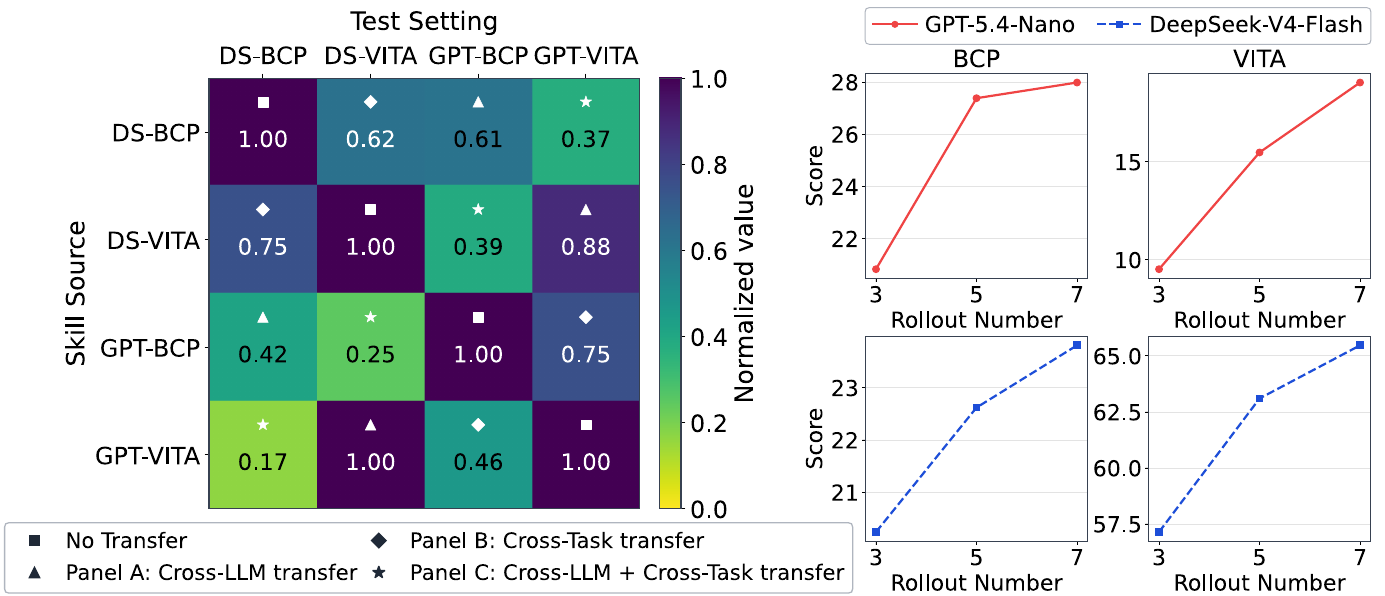} 
    \caption{\textbf{Left}: Skill transferability heatmap across LLMs (DS: DeepSeek-V4-Flash, GPT: GPT-5.4-Nano) and tasks (BCP: BrowseComp-Plus, VITA: VitaBench). \textbf{Right}: Performance scaling across increasing multi-trajectory rollout numbers ($K=3, 5, 7$).}
    \label{fig:ablation}
\end{figure}

To evaluate the generalization capabilities of Skill-MAS, we further analyze the transferability of Meta-Skills across different LLMs and domains. As shown in Table~\ref{tab:transfer_three_panels} and Figure~\ref{fig:ablation}~\textbf{Left}, \textbf{Skill Source} denotes the LLM-task pair used to evolve the Meta-Skill, and \textbf{Test Setting} denotes the pair used for evaluation. Table~\ref{tab:transfer_three_panels} details the absolute scores and the performance gains ($\Delta$) over ``Skill-MAS-init'', while Figure~\ref{fig:ablation} further visualizes these gains by column-wise normalization (scaling the $\Delta$ values from 0 to 1 for each Test Setting), where darker colors signify greater relative improvements.

The \textbf{No Transfer} scenario (matching Source and Test settings) achieves the maximum performance gains, dominating the diagonal of the heatmap. This is followed by \textbf{Panel A} (same task, different LLM), which aligns well with intuition: since the underlying task remains identical, analyzing and refining the trajectories naturally yields similar and highly transferable Meta-Skills, regardless of the specific LLM. Interestingly, \textbf{Panel B} (different task, same LLM) also delivers competitive performance. This robust out-of-domain generalization validates a core design choice in our evolution process: by explicitly prompting the Meta-agent to avoid domain-specific tricks and focus on general patterns, the optimized Meta-Skills successfully learn task-agnostic strategies. Therefore, they maintain effectiveness even on unseen datasets. Lastly, \textbf{Panel C} (different task, different LLM) exhibits the weakest performance, logically reflecting the extreme challenge of simultaneously transferring across both LLM and task distributions.

\subsubsection{Ablation on Multi-Trajectory Rollout}

In this section, we conduct an ablation study to analyze the impact of a key hyperparameter: the rollout number per sample during \textit{Multi-Trajectory Rollout}. As illustrated in Figure~\ref{fig:ablation}~\textbf{Right}, we evolve and evaluate Skill-MAS performance using rollout numbers of 3 and 7 (the default value is 5). Overall, the results indicate a clear positive correlation, where the performance consistently improves as the number of sampled trajectories increases. However, we observe a phenomenon of diminishing returns, which means the performance gain achieved by increasing the rollout number from 3 to 5 is larger than the gain from 5 to 7. And a higher rollout number inevitably incurs greater computational costs during the evolution phase. Therefore, in practical application, setting this parameter requires a careful trade-off between maximizing performance and managing evolution overhead.

\begin{table}[t]
    \centering
    \resizebox{0.48\textwidth}{!}{%
    \begin{tabular}{l *{4}{>{\centering\arraybackslash}m{1cm}}}
    \toprule
    \multirow{2.5}{*}{\textbf{Variants}} 
      & \multicolumn{2}{c}{\makebox[3cm][c]{\textbf{GPT-5.4-Nano}}}
      & \multicolumn{2}{c}{\makebox[3cm][c]{\textbf{DeepSeek-V4-Flash}}} \\
    \cmidrule(lr){2-3} \cmidrule(lr){4-5}
      & BCP & VITA & BCP & VITA \\ 
    \midrule
    Ours                & 27.38 & 15.48 & 22.62 & 63.10 \\
    Full-Validation     & 22.02 & 13.10 & 19.64 & 59.52 \\
    Half-Validation     & 21.43 & 9.52 & 17.26 & 58.33 \\
    Multi-task Learning & 20.83 & 16.67 & 22.02 & 64.29 \\
    \bottomrule
    \end{tabular}
    }
    \caption{Results of Skill-MAS with different settings.}
    \label{tab:multi_task_wo_label}
\end{table}

\subsubsection{Ablation on Selective Reflection}

Here, we conduct an ablation to examine the impact of label dependency during \textit{Selective Reflection}. By default, Skill-MAS uses ground-truth labels to calculate trajectory scores and prioritize samples. Since real-world data may lack labels, we disable this adaptive selection mechanism and test two label-free variants: \textbf{Full-Validation} (selecting all samples) and \textbf{Half-Validation} (randomly selecting a 50\% subset). 
Table~\ref{tab:multi_task_wo_label} demonstrates that without adaptive priority selection, while both variants suffer a performance drop, they surpass most baselines. On the one hand, this effectively highlights the crucial role of our adaptive priority selection. However, on the other hand, it indicates that Skill-MAS achieves suboptimal results in label-free settings. Therefore, an important avenue for future work is incorporating label-free components into skill evolution, such as utilizing the Meta-agent's self-confidence scores. By decoupling the Meta-Skill evolution from the requirement of ground-truth labels, we can significantly boost the framework's practicality for real-world deployment.

\subsubsection{Multi-Task Learning}

In addition to our primary single-domain setup (Table~\ref{tab:main_results}), we conduct an ablation study to explore a \textbf{Multi-task Learning} scenario. In this setting, the Meta-Skill is evolved on an aggregated pool of all four datasets before being evaluated on the corresponding test sets. Table~\ref{tab:multi_task_wo_label} shows that compared to our default setup, the multi-task variant achieves slight improvements on VitaBench, but performance drops on BrowseComp-Plus. This mixed result is not surprising, since Skill-MAS is not explicitly optimized for multi-task learning. Without mechanisms to adaptively analyze or isolate samples across different domains, the system may struggle to extract shared improvement patterns while ignoring domain-specific noise. Therefore, while the multi-task setting shows inherent promise, it requires more principled multi-task algorithms to be fully effective.

\subsection{Skill Evolution}

\begin{figure}[t]
    \centering
    \includegraphics[width=\linewidth]{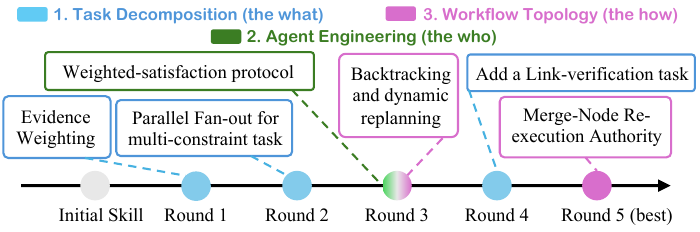} 
    \caption{Meta-Skill Evolution on BrowseComp-Plus.}
    \label{fig:evolution}
\end{figure}

Within the trajectory in Figure~\ref{fig:evolution} (DeepSeek-V4-Flash, BrowseComp-Plus), skill evolution traces a coherent arc from how evidence is framed, through how it is adjudicated, to how failures are remediated. Module~1 first establishes the epistemic scaffolding: constraint prioritization and fan-out retrieval transform decomposition from a flat partitioning of subtasks into a relational search plan with structural cues. On this foundation, Module~2 displaces brittle binary checks with calibrated evaluation, i.e., weighted constraint satisfaction with partial-evidence fallback. Finally, Module~3 improves the orchestration capability, where cross-entity bridging and merge-node re-execution shift the integration stage from passive aggregation to active evidence recovery. Taken together, these updates delineate a progressive Meta-Skill evolution from decomposition design, through agent-level epistemic control, to system-level resilience.

\section{Conclusion}
\label{sec:conclusion}

In this paper, we propose Skill-MAS, a novel framework that conceptualizes multi-agent orchestration as an evolvable Meta-Skill. By decoupling experiential learning from parametric updates, our approach enables frozen frontier LLMs to iteratively refine their architectural strategies through Multi-Trjectory Rollout and Selective Reflection. Extensive evaluations demonstrate that Skill-MAS achieves a better cost-performance trade-off and distills highly transferable orchestration principles across diverse domains and LLMs. Moving forward, by harnessing the experience distillation and excellent transferability of Meta-Skill, Skill-MAS unlocks new possibilities for deploying adaptive multi-agent systems in complex environments.

\section*{Limitations}


While Skill-MAS exhibits robust performance under supervised conditions with access to ground-truth labels, its effectiveness may degrade under weakly-supervised or unsupervised settings where such high-quality feedback is absent. To mitigate this, future work could design self-supervised evaluation mechanisms, such as integrating LLM-as-a-judge frameworks to guide the selective reflection phase without relying on external labels.

\section*{Acknowledgement}

This work is supported by Ant Group through CCF-Ant Research Fund (CCF-AFSG RF20250502).

\bibliography{custom}

\appendix

\section{Description of Appendix}
\label{appendix:description of appendix}

The appendix offers extended methodological details and experimental evidence to further substantiate the findings in the main text. \textbf{Appendix~\ref{appendix:pseudocode}} presents detailed pseudocode that illustrates the algorithmic workflow of the proposed Skill-MAS. \textbf{Appendix~\ref{appendix:statistics of benchmarks}} reports exhaustive benchmark statistics and descriptive summaries, including dataset splitting protocols and the characteristics of each domain-specific task. \textbf{Appendix~\ref{appendix:experimental details}} describes the full experimental setup, covering baselines and implementation details, and more ablation studies. \textbf{Appendix~\ref{appendix:case study}} provides an in-depth case study of the Meta-Skill and the generated MAS. \textbf{Appendix~\ref{appendix:prompt details}} compiles the complete set of prompts used in Skill-MAS and our experiments.

\section{Pseudocode of Skill-MAS}
\label{appendix:pseudocode}

\begin{algorithm}[H] 
\caption{Skill-MAS Evolution}
\label{alg:skill_mas}
\begin{algorithmic}[1]
\REQUIRE Validation set $\mathcal{T} = \{t_1, \dots, t_N\}$, Initial Skill $\mathcal{S}^{(1)}$, Max rounds $R$, Rollout size $K$
\ENSURE Optimal Meta-Skill $\mathcal{S}^*$

\STATE Initialize $S^* \leftarrow -1$, $\mathcal{S}^* \leftarrow \mathcal{S}^{(1)}$

\FOR{$r = 1$ to $R$}
    \vspace{6pt}
    \STATE \textbf{\textit{Stage 1: Multi-Trajectory Rollout}}
    \STATE $\mathcal{D}^{(r)} \leftarrow \emptyset$
    \FOR{each task $t_i \in \mathcal{T}$}
        \STATE Sample $K$ trajectories under current $\mathcal{S}^{(r)}$
        \STATE Record $\tau_{i,k} = (\mathrm{id}_i, k, s_{i,k}, \mathcal{S}^{(r)}, \Phi_{i,k})$
        \STATE $\mathcal{D}^{(r)} \leftarrow \mathcal{D}^{(r)} \cup \{\tau_{i,k}\}_{k=1}^K$
        \STATE Compute difficulty $d_i$ and uncertainty $u_i$
    \ENDFOR
    
    \STATE Calculate validation score $S^{(r)} = \frac{1}{N}\sum_i \bar{s}_i$
    \IF{$S^{(r)} > S^*$}
        \STATE Update $S^* \leftarrow S^{(r)}$ and $\mathcal{S}^* \leftarrow \mathcal{S}^{(r)}$
    \ENDIF

    \vspace{6pt}
    \textbf{\textit{Stage 2: Selective Reflection}}
    \STATE \textit{\# Priority-Driven Task Selection}
    \STATE Normalize $u_i \rightarrow \tilde{u}_i$ and $d_i \rightarrow \tilde{d}_i$ across $\mathcal{T}$
    \STATE Compute unified priority $p_i = \frac{1}{2}(\tilde{u}_i + \tilde{d}_i)$
    \STATE Sort task priority: $p_{(1)} \geq \dots \geq p_{(N)}$
    \STATE Compute 1st-order diff. $\delta_j = p_{(j)} - p_{(j+1)}$
    \STATE Find elbow $j^* = \arg\max_j |\delta_j - \delta_{j+1}|$
    \STATE Select target subset $\mathcal{T}_{\mathrm{sel}} = \{t_{(1)}, \dots, t_{(j^*)}\}$

    \STATE \textit{\# Hierarchical Trajectory Reflection}
    \FOR{each task $t_i \in \mathcal{T}_{\mathrm{sel}}$}
        \STATE Split trajectories into $\mathcal{H}_i$ and $\mathcal{L}_i$
        \STATE Contrastive diagnosis between $\mathcal{H}_i$ and $\mathcal{L}_i$ within-task and cross-task
        \STATE Generate report $\mathcal{R}_i$ and patch $\{\hat{\delta}_i\}$
    \ENDFOR
    \STATE Synthesize $\{\mathcal{R}_i\}$ to find systemic patterns
    \STATE Rank $\{\hat{\delta}_i\}$ to form structured evidence $\mathcal{E}$

    \textit{\# Skill Optimization}
    \STATE Update modules (Decomposition, Engineering, Orchestration) based on $\mathcal{S}^{(r)}$ and $\mathcal{E}$
    \STATE Abstract changes into principles
    \STATE Output revised skill $\mathcal{S}^{(r+1)}$
\ENDFOR

\RETURN $\mathcal{S}^*$
\end{algorithmic}
\end{algorithm}

\section{Statistics of Benchmarks}
\label{appendix:statistics of benchmarks}

Since the optimization of certain automatic-MAS relies on a validation set to discover the best MAS, we randomly sample examples to construct validation and test sets. For the sake of fair evaluation, all reported metrics are strictly based on the test set. The exact composition of these splits can be found in Table~\ref{tab:split of dataset}. Note that for Multi-task Learning, we select half of the validation set from each dataset to build the aggregate validation pool while evaluating on the corresponding test set.

\textbf{DeepResearchBench}~\cite{du2025deepresearch}: This benchmark aims to systematically evaluate the capabilities of agents in an autonomous research report writing. It consists of 100 meticulously crafted PhD-level research tasks across 22 distinct fields. The evaluation focuses on assessing four key dimensions (i.e., comprehensiveness, insight, instruction-following, and readability), serving as a rigorous test for deep research capabilities.

\textbf{Humanity's Last Exam-Math}~\cite{phan2025humanity}: This benchmark is designed to track the rapid advancements in LLM capabilities at the frontier of human knowledge. It encompasses 2,500 expert-level questions across dozens of subjects, focusing on complex mathematical and scientific reasoning. This benchmark rigorously assesses the model's accuracy and calibration on tasks that push the limits of expert human capabilities. We select the MATH subset and conduct experiments on such complex mathematical questions. We randomly sample 200 questions for our experiment.

\textbf{BrowseComp-Plus}~\cite{chen2025browsecomp}: This benchmark evaluates agents' capability on complex multi-hop queries that require iterative search planning and reasoning. To ensure fair and transparent evaluations, it employs a fixed, carefully curated corpus equipped with human-verified supporting documents and challenging negative samples. Evaluation is based on accuracy, enabling well-controlled experiments to test models' dynamic reasoning capabilities. We randomly sample 200 questions for our experiment.

\textbf{VitaBench}~\cite{he2025vitabench}: This benchmark evaluates language agents on versatile interactive tasks grounded in real-world daily scenarios, such as food delivery, in-store consumption, and online travel services. It features a complex life-serving simulation environment comprising 66 tools, encompassing 100 cross-scenario tasks and 300 single-scenario tasks. Utilizing a rubric-based sliding window evaluator, the benchmark rigorously measures an agent's success rate in navigating dynamic user interactions, reasoning across temporal and spatial dimensions, proactively clarifying ambiguous instructions, and utilizing complex tool sets.
In our evaluation, we adopt the cross-scenario tasks to test models' ability.

\begin{table*}[htbp]
    \centering
    \resizebox{\textwidth}{!}{%
    \begin{tabular}{@{}lccccc@{}}
    \toprule
    Split  & DeepResearchBench & Humanity’s Last Exam-Math & BrowseComp-Plus & VitaBench & Multi-task Learning \\ \midrule
    Validation & 16 & 32 & 32 & 16 & 48 \\ 
    Test & 84 & 168 & 168 & 84 & - \\
    \bottomrule
    \end{tabular}%
    }

    \caption{Data size for each split in each dataset.}
    \label{tab:split of dataset}
\end{table*}

\begin{table*}[htbp]
    \centering
    \begin{tabular}{@{}llcc@{}}
    \toprule
    Category & Hyperparameter  & Description & Value \\ \midrule
    \multirow{2}{*}{Skill-MAS}& $R$ & The round number of skill evolution & 10 \\
    & $K$ & The rollout number of each task & 5 \\
    \midrule
    \multirow{2}{*}{LLM calls}& temperature & The sampling temperature of calling LLM & 1.0 \\
    & max\_tokens & The maximum number of output tokens & 32768 \\
    \bottomrule
    \end{tabular}%

    \caption{The description and value of important hyperparameters.}
    \label{tab:hyper-parameters}
\end{table*}

\section{Experimental Details}
\label{appendix:experimental details}

\subsection{Automatic MAS Baselines}
\label{appendix:automatic mas baselines}

\textbf{EvoAgent}~\cite{yuan2025evoagent}: 
Inspired by evolutionary algorithms, EvoAgent dynamically expands specialized single agents into multi-agent configurations. By applying evolutionary operators such as mutation, it autonomously spawns agents with diverse settings to tackle complex tasks in real time.

\textbf{AOrchestra}~\cite{ruan2026aorchestra}: 
AOrchestra adopts a dynamic orchestration paradigm where a central orchestrator instantiates tailored sub-agents on demand. It operates over a unified abstraction, continuously curating task-relevant contexts and delegating execution to dynamically created agents equipped with specific tools and models.

\textbf{AFlow}~\cite{zhang2024aflow}: 
AFlow treats workflow optimization as an automated search problem over code-represented reasoning chains. Utilizing techniques like Monte Carlo Tree Search, it iteratively refines agentic workflows based on execution feedback at inference time.

\textbf{MAS$^2$}~\cite{wang2025mas}: 
MAS$^2$ shifts from a rigid ``generate-once-and-deploy'' paradigm to a recursive self-generation approach. Leveraging a dedicated tri-agent team during an offline training phase, it systematically searches for and solidifies optimal agent topologies, yielding highly customized and ready-to-deploy multi-agent systems for specific domains.

\textbf{MAS-Orchestra}~\cite{ke2026mas}: 
MAS-Orchestra formulates multi-agent orchestration as a function-calling reinforcement learning problem optimized at train time. By abstracting complex sub-agents as callable functions, it learns a holistic orchestration policy offline, enabling the central orchestrator to generate a complete and highly optimized MAS architecture in a single decision step.

\subsection{Implementation Details}
\label{appendix: implementation details}
To balance optimization quality with computational expenditure, we restrict AFlow's maximum search iterations to 10 and evaluate the validation set three times per iteration. For all other baselines, we strictly follow the original settings. The core hyperparameters adopted for our Skill-MAS are summarized in Table~\ref{tab:hyper-parameters}. When configuring the backbone LLMs, we set GPT-5.4-Nano and Gemini-3.1-Flash with a ``low'' reasoning effort. Conversely, for Qwen3.5-Plus and DeepSeek-V4-Flash, we utilize their standard versions without additional reasoning effort overhead. We deploy Gemini-3.1-Flash as the default LLM-judge.

\subsection{Cost Analysis}
\label{appendix:cost analysis}

As discussed in the main text, Training-time MAS relies on GPU resources during training, whereas some inference-time MAS and our Skill-MAS require a validation set to iteratively generate the final MAS or evolve the Meta-Skill. On the one hand, the GPU and token costs associated with these training or evolution phases are difficult to align for a direct comparison. On the other hand, once the orchestrator is trained or the Meta-Skill is optimized, it can be reused to generate MAS for different input queries during the inference stage, where the inference cost is typically the bottleneck for real-world deployment. Therefore, the training/evolution costs are excluded from the main text, and the reported costs in Table~\ref{tab:main_results} refer strictly to the inference overhead on the test set. 

\emnlpcr{
Here, we provide the evolution costs of Skill-MAS on the validation set in Table~\ref{tab:Skill-MAS cost}.
Since the evolution cost is a one-time offline cost and the optimized Meta-Skill can be reused, while inference-time MAS search is paid repeatedly for each test query. 
We furthermore add an amortized analysis:
$
C\_amortized(Q) = C\_infer + C\_evolve / Q
$
where Q is the number of tasks or deployments that reuse the same Meta-Skill. As a concrete example in the DeepSeek-V4-flash setting, Table~\ref{tab:Skill-MAS cost} reports an evolution cost of 24.54 USD. The inference cost of Skill-MAS is 2.12 USD, while the EvoAgent baseline costs 3.33 USD. The break-even reuse count is therefore 24.54 / (3.33 - 2.12) = 20.28, i.e., about 21 reused queries or deployments. After this point, the one-time evolution cost is recovered, and Skill-MAS becomes increasingly cost-effective in repeated-query settings.
}

\emnlpcr{
Furthermore, Table~\ref{tab:cost gain} shows the cost-gain tradeoff.
First, we report the cost-performance trade-off under different multi-trajectory rollout budgets (K=3, 5, 7). Increasing K provides more diverse trajectories and generally improves the quality of reflection, but it also increases the rollout cost. The results show a clear trade-off: gains increase with larger K, while the marginal improvement gradually saturates.
Second, we compare Selective Reflection with a full-validation reflection variant. Since Selective Reflection only sends high-priority samples into the reflection stage, it reduces the number of trajectories that require expensive LLM-based analysis. This gives lower evolution cost than reflecting over the full validation set.
}

\begin{table}[htbp]
    \centering
    \begin{tabular}{@{}lcc@{}}
    \toprule
    Variant & Cost & Acc \\\midrule
    K=3 & 13.63 & 20.24 \\ 
    K=5 (Ours) & 21.75 & 22.62 \\ 
    K=7 & \textbf{31.18} & \textbf{23.81} \\ 
    \midrule
    Full validation reflection & \textbf{23.44} & 19.64 \\ 
    Selective Reflection (Ours) & 21.75 & \textbf{22.62} \\ 
    \bottomrule
    \end{tabular}%

    \caption{Cost-gain studies on BrowseComp-plus with DeepSeek-V4-Flash.}
    \label{tab:cost gain}
\end{table}

\subsection{More Ablation Studies}
\emnlpcr{
Table~\ref{tab:more ablation} provides more ablation studies as supplementary experiments to demonstrate the effectiveness of Skill-MAS.
First, we add an ablation variant (removing cross-task analysis), and the result shows that it pushes the reflection process toward case-specific fixes and overfitting to individual examples.
Second, we add two additional skill-evolution baselines, both initialized from the same Skill-MAS-init: (1) Skill-MAS-init + prompt optimization, following the idea of reflective prompt evolution in GEPA~\cite{agrawal2026gepa}; and (2) Skill-MAS-init + self-improving agent, following the experience-driven self-evolution setting in EvolveR~\cite{wu2025evolver}. The overall comparison shows that the two generic skill-evolution baselines improve over the initial skill but underperform Skill-MAS-optimized. This supports that the main gain is not merely due to ordinary prompt optimization or experience reuse, but to the MAS-specific evolution of orchestration knowledge.
}

\begin{table}[htbp]
    \centering
    \begin{tabular}{@{}lc@{}}
    \toprule
    Variant & Acc \\\midrule
    Skill-MAS-optimized & \textbf{22.62} \\ 
    + removing cross-task analysis & 17.86 \\ \midrule
    Skill-MAS-init & 15.48 \\ 
    + GEPA & 19.05 \\ 
    + EvolveR & 18.45 \\ 
    \bottomrule
    \end{tabular}%

    \caption{More ablation studies on BrowseComp-plus with DeepSeek-V4-Flash.}
    \label{tab:more ablation}
\end{table}

\begin{table*}[htbp]
    \centering
    \begin{tabular}{@{}cccc@{}}
    \toprule
    Gemini-3.1-Flash & GPT-5.4-Nano & Qwen3.5-Plus & DeepSeek-V4-Flash \\\midrule
    9.35 & 31.36 & 59.06 & 24.54\\ 
    \bottomrule
    \end{tabular}%

    \caption{Average cost (USD \$) of Skill-MAS on four benchmarks using different Meta-agents.}
    \label{tab:Skill-MAS cost}
\end{table*}

\section{Case Study}
\label{appendix:case study}

We show the initial Meta-Skill in Figure~\ref{fig:init_skill} and the four optimized Meta-Skills of the corresponding benchmark from Figure~\ref{fig:skill_drb_1} to Figure~\ref{fig:skill_vita_2}.

Compared to the initial skill, the optimized skills move from a generic framework to a much more operational specification. They introduce explicit structural constraints for decomposition and orchestration (e.g., bounded parallelism, dedicated merge/synthesis stages, capability-boundary splitting), together with formal decision and validation rules. As a result, agent behavior becomes less ambiguous, and the overall MAS is better aligned with complex reasoning tasks.

Across datasets, a consistent pattern is the addition of reliability-oriented controls: constraint-aware reasoning, structured output contracts, verification gates, and backtracking mechanisms. These shared designs reduce common failure modes such as premature commitment, error propagation across stages, and unstable downstream handoff. Although each dataset emphasizes different aspects (e.g., interpretation calibration for math and evidence weighting for retrieval), they converge on the same advantage: a more robust, auditable, and transferable orchestration policy.

We compare the generated MAS of Skill-MAS and baselines in Table~\ref{tab:mas-compare-bcp} and Table~\ref{tab:mas-compare-vita}.
Across both case studies, Skill-MAS-optimized improves over baseline automatic MAS by replacing repeated or loosely coordinated search with structured decomposition and staged verification. In BrowseComp-Plus, it splits the query into clue-specific retrieval branches and enforces cross-clue consistency before the final decision, so the answer is supported by linked evidence rather than a single search path.
Compared with Skill-MAS-init, the key gain is moving from a linear pipeline to a branched workflow. In VitaBench, explicit ``explore, evaluate, order'' branches for meal, book bar, and train tasks reduce error propagation and preserve constraint fidelity, yielding more robust outcomes.

\begin{figure*}[htbp]
    \centering
    \begin{tcolorbox}[
        enhanced,
        colframe=Salmon!90!Black,
        colback=Salmon!20,
        coltitle=white,
        fonttitle=\large\bfseries,
        title={The initial Meta-Skill},
        halign title=left,
        fontupper=\footnotesize,
        boxrule=1pt,
        arc=3mm,
        boxsep=2pt,
        left=8pt,
        right=8pt,
        top=4pt,
        bottom=4pt
    ]

\hspace{0em}---

\hspace{0em}name: unified\_meta\_agent\_skill

\hspace{0em}description: "A foundational meta-agent skill for generating Multi-Agent Systems (MAS). It systematically drives the process from conceptual task decomposition to agent engineering, and finally to workflow orchestration."

\hspace{0em}tags:

\hspace{2em}- meta-agent

\hspace{2em}- task-decomposition

\hspace{2em}- agent-engineering

\hspace{2em}- workflow-orchestration

\hspace{0em}inputs:

\hspace{2em}- user\_query

\hspace{0em}---

\vspace{6pt}
\hspace{0em}\textbf{1. Task Decomposition Module (The "What")}

\hspace{0em}Core Objective: Analyze the user query and break it down into a logical blueprint.

\vspace{3pt}
\hspace{0em}- Intent \& Scope Analysis: Understand the macro objective, identify core requirements, and define the boundaries of the task.

\hspace{0em}- Sub-task Breakdown: Decompose the high-level request into a set of discrete, manageable, and logically cohesive sub-tasks.

\hspace{0em}- Logical Dependency Mapping: Identify the business-logic relationships between sub-tasks (e.g., prerequisite, parallel, or iterative). Note: This focuses on logical order, not system dataflow.

\hspace{0em}- Success Criteria: Define clear objective outcomes for each sub-task to ensure evaluability.

\vspace{6pt}
\hspace{0em}\textbf{2. Agent Engineering Module (The "Who")}

\hspace{0em}Core Objective: Design specialized sub-agents tailored for the sub-tasks defined in Stage 1.

\vspace{3pt}
\hspace{0em}- Role Profiling: Assign a unique identity and specialized role to each sub-agent based on its target sub-task.

\hspace{0em}- Instruction Design: Draft precise system prompts/instructions. Define the agent's specific goals, behavioral boundaries, and output expectations.

\hspace{0em}- Input Context Framing: Specify what contextual information this agent requires from the user or the global task to begin its work.

\vspace{6pt}
\hspace{0em}\textbf{3. Workflow \& Orchestration Module (The "How")}

\hspace{0em}Core Objective: Wire the distinct agents from Stage 2 into a functional, executable Multi-Agent System (MAS).

\vspace{3pt}
\hspace{0em}- Architectural Topology: You can design the optimal MAS architecture (e.g., Sequential Pipeline, Router-based, Hierarchical, or Blackboard) based on Stage 1's logical dependencies. For those complex but important sub-tasks, you can design a localized, different topology design using the instantiated agents. For example, you can define an iterative loop for those sub-tasks that need a cycle check to ensure high quality. Or you can call the same agent multiple times to generate diverse outputs and give all of them to the following sub-agents.

\hspace{0em}- Dataflow \& State Management: Define the exact I/O mapping. Specify how the output schema of one agent transforms into the input payload/prompt of downstream agents. Determine how the global context (memory/state) is maintained.

\hspace{0em}- Executable Generation: Output the final orchestration logic/code structure that binds the agents, tools, and dataflow into a ready-to-run system.

    \end{tcolorbox}
    \caption{Illustration of the initial Meta-Skill used for Skill-MAS-init and Skill-MAS evolution.}
    \label{fig:init_skill}
\end{figure*}

\begin{figure*}[htbp]
    \centering
    \begin{tcolorbox}[
        enhanced,
        colframe=Salmon!90!Black,
        colback=Salmon!20,
        coltitle=white,
        fonttitle=\large\bfseries,
        title={Optimized Meta-Skill for DeepResearchBench (DeepSeek-V4-Flash, Part 1/3)},
        halign title=left,
        fontupper=\footnotesize,
        boxrule=1pt,
        arc=3mm,
        boxsep=2pt,
        left=8pt,
        right=8pt,
        top=4pt,
        bottom=4pt
    ]

\hspace{0em}\textbf{1. Task Decomposition Module (The "What")}

\hspace{0em}Core Objective: Analyze the user query and break it down into a logical blueprint.

\vspace{3pt}
\hspace{0em}- Intent \& Scope Analysis: Understand the macro objective, identify core requirements, and define the boundaries of the task.

\hspace{0em}- Sub-task Breakdown: Decompose the high-level request into a set of discrete, manageable, and logically cohesive sub-tasks.

\hspace{0em}\hspace{2em}- Structural Topology Enforcement: For any task whose deliverable requires integrating multiple analytical components into a coherent final synthesis (e.g., a research report, comparative analysis, multi-source case study), the decomposition MUST enforce the following canonical topology:

\hspace{0em}\hspace{4em}1. Context-Scoping Root Node: A dedicated root sub-task that defines scope, key concepts, metrics, terminology, and evaluation criteria before any analytical work begins. This node frames the entire problem space and ensures all downstream agents operate under a shared understanding.

\hspace{0em}\hspace{4em}2. Parallel Analytical Branches: One sub-task per distinct analytical component (capped at four branches). These must be designed to run in parallel from the context-scoping root, with no intermediate sequential dependencies among them.

\hspace{0em}\hspace{4em}3. Dedicated Synthesis Terminal Node: A final sub-task that receives the outputs of all parallel branches and integrates them into the requested cohesive output (e.g., report, article, synthesis). The synthesis node must be the only terminal node.

\hspace{0em}\hspace{2em}- Hard Constraint: Strict sequential chaining of analytical components is disallowed for such tasks. If the query describes a chain where two or more downstream components depend on the same upstream prerequisite, the upstream node must fan‑out to those downstream components in parallel. The maximum depth from root to synthesis must be limited to two layers (root → parallel → synthesis). This structural constraint prevents cumulative error propagation and ensures each analytical branch receives equal context.

\hspace{0em}\hspace{2em}- Capability-Abstraction and Dependency Constraint: All analytical sub-task descriptions MUST be phrased as generic, capability-oriented scopes (e.g., "Capability to analyze methods for...") rather than concrete method-level or domain-specific actions. The dependency graph must contain edges only from the root to analytical branches and from branches to the synthesis node; no cross-branch edges are permitted. For multi-component or multi-vehicle tasks, each distinct sub-aspect must correspond to its own orthogonal branch (minimum three if the task naturally decomposes into three or more independent perspectives). This abstraction prevents narrow, overlapping outputs and ensures each branch can flexibly cover its assigned domain.

\hspace{0em}- Final Integrative Step Mandate: If the user query implies a final deliverable that synthesizes outputs from multiple sub-tasks (e.g., a report, article, analysis), a dedicated synthesis sub-task MUST be defined as the terminal node. For tasks that involve integration of distinct analytical streams, the context-scoping root node is also mandatory. This ensures the MAS produces a cohesive, integrated output rather than raw intermediate results.

\hspace{0em}- Logical Dependency Mapping: Identify the business-logic relationships between sub-tasks (e.g., prerequisite, parallel, or iterative). Note: This focuses on logical order, not system dataflow.

\hspace{0em}- Token-Budget \& Topology Planning: Estimate the token consumption of each sub-task and design the overall pipeline topology to prevent context window overflow. For multi-component synthesis tasks (where independent analytical sub-tasks feed into a final integration), prefer a parallel fan-in architecture (flat DAG) over a strict sequential pipeline to avoid token starvation of downstream agents. Consider merging prerequisite steps (e.g., data preparation) into the first analytical sub-task when they are not the main analytical goal, reducing node count and saving token capacity.

\hspace{0em}\hspace{2em}- Resource Budgeting and Constraint Validation: Derive a maximum sub-task count and per-sub-task output length limit such that the sum of all expected outputs fits within the total budget. Simultaneously, enforce a hard minimum resource floor: each analytical sub-task must be allocated at least 4000 expected tokens, each synthesis sub-task at least 6000 tokens, and the total across all sub-tasks must be at least 20000 tokens. Additionally, enforce that no analytical sub-agent receives less than 3000 characters (or equivalent tokens) to prevent starvation. If the proposed decomposition violates either the upper constraint (overflow) or the lower constraint (under-allocation), reject the plan and regenerate a more conservative or more generous allocation accordingly. When scaling to meet the minimum, increase per-sub-task limits proportionally (capping individual limits at a reasonable maximum). This pre‑commitment prevents both over‑ambitious planning that starves downstream agents and under‑resourced plans that cause cascade failure.

\hspace{0em}\hspace{2em}- Output Specification Contract per Sub-task: As part of resource budgeting, each sub-task definition must include an explicit output specification contract specifying: (a) a target token budget, (b) a list of representative domain terms that the sub-task output should cover (derived from its capability-oriented description), and (c) a minimum structural complexity (e.g., required number of sections, bullet points, or tables). This contract forms the basis for agent self-validation in Stage 2 and ensures that downstream agents receive consistent, quality-assured inputs. The contract must be passed to Stage 2 agent definition, where it is automatically injected into each agent's role instruction.

\hspace{0em}- Success Criteria: Define clear objective outcomes for each sub-task to ensure evaluability.

    \end{tcolorbox}
    \caption{Illustration of the optimized Meta-Skill for DeepResearchBench (DeepSeek-V4-Flash, Part 1/3).}
    \label{fig:skill_drb_1}
\end{figure*}

\begin{figure*}[htbp]
    \centering
    \begin{tcolorbox}[
        enhanced,
        colframe=Salmon!90!Black,
        colback=Salmon!20,
        coltitle=white,
        fonttitle=\large\bfseries,
        title={Optimized Meta-Skill for DeepResearchBench (DeepSeek-V4-Flash, Part 2/3)},
        halign title=left,
        fontupper=\footnotesize,
        boxrule=1pt,
        arc=3mm,
        boxsep=2pt,
        left=8pt,
        right=8pt,
        top=4pt,
        bottom=4pt
    ]

\hspace{0em}\textbf{2. Agent Engineering Module (The "Who")}

\hspace{0em}Core Objective: Design specialized sub-agents tailored for the sub-tasks defined in Stage 1.

\vspace{3pt}
\hspace{0em}- Role Profiling: Assign a unique identity and specialized role to each sub-agent based on its target sub-task.

\hspace{0em}- Instruction Design: Draft precise system prompts/instructions. Define the agent's specific goals, behavioral boundaries, and output expectations.

\hspace{0em}\hspace{2em}- Data Completeness Mandate: For any agent that must collect or estimate data, explicitly instruct that every requested field must be populated. If exact values are unavailable, the agent must provide plausible estimated values (e.g., using \~{} prefix). Prohibit returning placeholders like 'Data not available' or 'Insufficient information', as such gaps cascade into downstream failures.

\hspace{0em}\hspace{2em}- Output Length Contract and Specification Injection: For each agent, define a strict maximum output length (e.g., in tokens or word count) and embed it within the system instruction. Additionally, the agent's role instruction must include an auto-generated output specification block derived from the Stage 1 per-sub-task contract: (a) target token budget, (b) list of required domain terms, (c) minimum structural complexity (e.g., section headers, bullet points). Mandate a structured output format that aligns with the structural requirements. Include self‑validation against this specification. This standardized injection ensures every agent has unambiguous, formal goals.

\hspace{0em}\hspace{2em}- Graduated Two-Tier Validation with Corrective Injection: Every agent instruction must contain a built-in self-validation check using a two-tier framework:

\hspace{0em}\hspace{4em}- Strict Pass: Output reaches >= 60\% of target token budget, covers all required domain terms, and meets structural complexity requirements. Proceed normally.

\hspace{0em}\hspace{4em}- Soft Pass: Output reaches between 40\% and 60\% of target token budget and covers at least one required domain term. Accept the output but append a corrective note (e.g., "Soft pass: coverage of \textless{}missing terms\textgreater{} and token count \textless{}X\%\textgreater{} below threshold. Please expand in synthesis."). This note is stored in global memory and read by the synthesis agent.

\hspace{0em}\hspace{4em}- Fallback: If after retries (see retry protocol) neither strict nor soft pass is achieved, use the longest non-empty output from all attempts and attach a soft pass note.

\hspace{0em}\hspace{4em}- Retry Protocol with Threshold Reduction: On validation failure, re-run the agent with a targeted expansion instruction specifying which criteria were missed. Cap retries at 2. After each retry, reduce the token threshold by 10\% (relative to the original budget) and reduce the required domain term count by one (floor of 1). Apply the same reduction schedule consistently across all agents. For synthesis agents, require three structural elements and three domain terms; apply analogous reductions. The retry prompt must explicitly state: "Your previous output met \textless{}X\%\textgreater{} of the token target and covered \textless{}Y\textgreater{} of \textless{}Z\textgreater{} required domain terms. Please expand to at least \textless{}adjusted\_token\_target\textgreater{} tokens, covering \textless{}adjusted\_terms\textgreater{}, and include structural elements like headers or bullet points."

\hspace{0em}\hspace{4em}- Language Consistency: Ensure that validation criteria, key terms, and failure messages are written in the same language as the task description to avoid false failures.

\hspace{0em}\hspace{2em}- Boundary Conditions and Escalation Protocols: For each agent, define explicit boundary conditions that trigger escalation or self‑correction. These include: (a) when input data is insufficient to produce a complete output, (b) when the agent's confidence in its analysis falls below a threshold (e.g., unable to verify key claims), (c) when the agent detects contradictions in the input or its own intermediate reasoning. In such cases, the agent must include a self‑assessment note within its output (e.g., “Note: this section relies on estimated data due to lack of verified sources”) rather than suppressing the output. Instruct agents to first attempt self‑correction by re‑examining available context and applying conservative estimates, and only escalate to a human-like oversight mechanism if the issue persists. This prevents silent failures and preserves information flow.

\hspace{0em}- Input Context Framing: Specify what contextual information this agent requires from the user or the global task to begin its work.

    \end{tcolorbox}
    \caption{Illustration of the optimized Meta-Skill for DeepResearchBench (DeepSeek-V4-Flash, Part 2/3).}
    \label{fig:skill_drb_2}
\end{figure*}

\begin{figure*}[htbp]
    \centering
    \begin{tcolorbox}[
        enhanced,
        colframe=Salmon!90!Black,
        colback=Salmon!20,
        coltitle=white,
        fonttitle=\large\bfseries,
        title={Optimized Meta-Skill for DeepResearchBench (DeepSeek-V4-Flash, Part 3/3)},
        halign title=left,
        fontupper=\footnotesize,
        boxrule=1pt,
        arc=3mm,
        boxsep=2pt,
        left=8pt,
        right=8pt,
        top=4pt,
        bottom=4pt
    ]

\hspace{0em}\textbf{3. Workflow \& Orchestration Module (The "How")}

\hspace{0em}Core Objective: Wire the distinct agents from Stage 2 into a functional, executable Multi-Agent System (MAS).

\vspace{3pt}
\hspace{0em}- Architectural Topology: Design the optimal MAS architecture (e.g., Sequential Pipeline, Router-based, Hierarchical, or Blackboard) based on Stage 1's logical dependencies. For complex sub-tasks, embed localized topology patterns such as iterative refinement loops or parallel invocation of the same agent to produce diverse outputs for subsequent synthesis. Always align topology with the resilience constraints below.

\hspace{0em}\hspace{2em}- Topological Constraint for Resilience: For tasks that involve multiple independent analytical perspectives feeding into a synthesis, prefer a diamond fan-in topology with a root node, parallel analytical branches, and a single synthesis node. Avoid intermediate integration nodes between parallel branches and the synthesis, as each extra layer introduces a failure point. The maximum depth from root to synthesis should be at most 2 layers. This constraint, enforced during Stage 1 parsing, reduces the risk of cascading collapse from deep dependency chains.

\hspace{0em}\hspace{2em}- Graduated Validation and Retry Resilience at Every Node: Every sub‑task node in the pipeline must be protected by a retry‑with‑context‑preservation loop using the two-tier validation system defined in Stage 2. After each agent execution, evaluate its output against the specification contract: deliberate on strict pass vs. soft pass criteria, execute the threshold reduction schedule on failure, and maintain a soft pass note for the global memory. If retries are exhausted without a strict pass, use the longest non-empty output (from all attempts) and attach a soft pass note as a corrective injection. This eliminates the binary failure mode where valid near-miss outputs are discarded.

\hspace{0em}- Dataflow \& State Management: Define the exact I/O mapping. Specify how the output schema of one agent transforms into the input payload/prompt of downstream agents. Determine how global context (memory/state) is maintained.

\hspace{0em}\hspace{2em}- Global vs. Local Memory Management: Implement two distinct memory layers to maintain consistency across the workflow. Global Memory holds task‑level shared knowledge: the problem statement, scope definitions, agreed‑upon metrics, key terms, and any intermediate results needed across multiple sub‑tasks. After each agent completes execution, its output must be distilled into a standardized compact summary and appended to Global Memory. The summary must follow a fixed template: each agent's entry consists of a one‑line agent identifier (e.g., \texttt{Agent\_RoleName:}) followed by 3–5 bullet points, each bullet point <=150 characters, total <=750 characters per agent. Use plain markdown bullet points without extraneous labels. This template ensures Global Memory remains lean. Additionally, append any soft pass notes generated by the two-tier validation to a dedicated section of Global Memory (e.g., \texttt{Soft Pass Notes:}). The synthesis agent must read both the summaries and the soft pass notes to understand coverage gaps. Local Memory is per‑agent working memory used exclusively for the current sub‑task; it is cleared after the agent's output is harvested. Agents must be instructed to consult Global Memory at the start of their execution to align with established context, and to write back a structured summary upon completion.

\hspace{0em}- Pre-Synthesis Coverage Check (Dimension Verification): Before triggering the synthesis agent, perform a lightweight validation that verifies each required analytical dimension (as defined in Stage 1's sub-task breakdown) is represented in the collected outputs. This check uses keyword matching against the domain terms from each branch's specification contract. If a dimension is missing or under-represented, inject a specific expansion prompt into the synthesis agent's context (e.g., "The following required dimension appears to have insufficient coverage: \textless{}dimension\textgreater{}. You must include a dedicated section addressing it."). This prevents incomplete synthesis.

\hspace{0em}- Validation Gate Insertion: For critical handoffs, insert a lightweight validation step that checks the downstream agent's input against the contract (schema, minimum content length, no placeholder tokens). If validation fails, trigger a regeneration loop or escalate to a human-like oversight mechanism.

\hspace{0em}- Graceful Degradation on Retry Exhaustion: If an agent fails to produce a valid output after exhausting retries, the system must not insert a generic placeholder. Instead, it must pass the last best attempt (truncated to the token budget if necessary) as the downstream input, preserving any substantive content it contains. This prevents the total pipeline collapse that previously occurred when empty fallbacks propagated. Use the longer of the two attempts if their token counts differ by more than 500; otherwise, merge both attempts to retain partial content from each.

\hspace{0em}- Executable Generation: Output the final orchestration logic/code structure that binds the agents, tools, and dataflow into a ready-to-run system.

    \end{tcolorbox}
    \caption{Illustration of the optimized Meta-Skill for DeepResearchBench (DeepSeek-V4-Flash, Part 3/3).}
    \label{fig:skill_drb_3}
\end{figure*}

\begin{figure*}[htbp]
    \centering
    \begin{tcolorbox}[
        enhanced,
        colframe=Salmon!90!Black,
        colback=Salmon!20,
        coltitle=white,
        fonttitle=\large\bfseries,
        title={Optimized Meta-Skill for HLE-MATH (DeepSeek-V4-Flash, Part 1/2)},
        halign title=left,
        fontupper=\footnotesize,
        boxrule=1pt,
        arc=3mm,
        boxsep=2pt,
        left=8pt,
        right=8pt,
        top=4pt,
        bottom=4pt
    ]

\hspace{0em}---

\hspace{0em}name: unified\_meta\_agent\_skill

\hspace{0em}description: "A foundational meta-agent skill for generating Multi-Agent Systems (MAS). It systematically drives the process from conceptual task decomposition to agent engineering, and finally to workflow orchestration."

\hspace{0em}tags:

\hspace{2em}- meta-agent

\hspace{2em}- task-decomposition

\hspace{2em}- agent-engineering

\hspace{2em}- workflow-orchestration

\hspace{0em}inputs:

\hspace{2em}- user\_query

\hspace{0em}---

\vspace{6pt}
\hspace{0em}\textbf{1. Task Decomposition Module (The "What")}

\hspace{0em}Core Objective: Analyze the user query and break it down into a logical blueprint.

\vspace{3pt}
\hspace{0em}- Intent \& Scope Analysis: Understand the macro objective, identify core requirements, and define the boundaries of the task.

\hspace{0em}- Sub-task Breakdown: Decompose the high-level request into a set of discrete, manageable, and logically cohesive sub-tasks.

\hspace{0em}- Logical Dependency Mapping: Identify the business-logic relationships between sub-tasks (e.g., prerequisite, parallel, or iterative). Note: This focuses on logical order, not system dataflow.

\hspace{0em}- Success Criteria: Define clear objective outcomes for each sub-task to ensure evaluability.

\hspace{0em}- Mandatory Interpretation Register (MIR) with Interpretation Propagation Mandate: Before finalizing the decomposition, systematically surface and document all implicit assumptions, ambiguous terms, and unstated constraints present in the query. For each ambiguous term, enumerate at least two plausible interpretations, evaluate their consistency with the query's context, and resolve by selecting the interpretation that is empirically validated. The output must be a formal, structured register (e.g., a specification document or contractual list) that is treated as a binding constraint by all downstream agents. Include a fallback instruction: if the selected interpretation leads to a trivial or contradictory result, flag the anomaly and attempt the next most plausible interpretation. This register shall be explicitly communicated to subsequent modules to eliminate divergent interpretations and prevent cascading failures from a single flawed premise. This step is non-optional: if no ambiguities are apparent, the register must still note that assumption of clear semantics was validated.

\hspace{0em}\hspace{2em}- Resolution Contract: The MIR must include a Resolution Contract that explicitly lists for each ambiguous term: (a) all candidate interpretations, (b) the chosen interpretation with formal justification, (c) the rejected interpretations and reasons for rejection, and (d) a fallback chain of alternative interpretations. This contract is immutable across all stages; any downstream agent that discovers a contradiction must halt and escalate to the planning agent for re-resolution.

\hspace{0em}\hspace{2em}- Small-n Calibration for Interpretation Validation (Mandatory): For every ambiguous term or construct, the MIR must include a calibration test on the smallest non-trivial instance (e.g., n=2,3,4 or equivalent). For each candidate interpretation, compute the expected outcome for that small instance using the interpretation. Compare against known ground truth or brute-force enumeration of that minimal case. The selected interpretation must be the one whose small-n prediction matches ground truth. If no interpretation passes, flag the ambiguity as unresolvable and either attempt fallback interpretations or escalate to a human-in-the-loop. The calibration result is embedded in the Resolution Contract and must be explicitly referenced by all downstream agents. A runtime adherence check at each stage boundary shall verify that the agent's output references the chosen interpretation and confirms consistency with the small-n calibration; if missing or contradictory, the agent is re-executed with strengthened instruction to comply. This ensures that the planner's semantic resolution is empirically grounded and binding throughout the pipeline.

\vspace{6pt}
\hspace{0em}\textbf{2. Agent Engineering Module (The "Who")}

\hspace{0em}Core Objective: Design specialized sub-agents tailored for the sub-tasks defined in Stage 1.

\vspace{3pt}
\hspace{0em}- Role Profiling: Assign a unique identity and specialized role to each sub-agent based on its target sub-task.

\hspace{0em}- Instruction Design: Draft precise system prompts/instructions. Define the agent's specific goals, behavioral boundaries, and output expectations.

\hspace{0em}- Input Context Framing: Specify what contextual information this agent requires from the user or the global task to begin its work.

    \end{tcolorbox}
    \caption{Illustration of the optimized Meta-Skill for HLE-MATH (DeepSeek-V4-Flash, Part 1/2).}
    \label{fig:skill_hlemath_1}
\end{figure*}

\begin{figure*}[htbp]
    \centering
    \begin{tcolorbox}[
        enhanced,
        colframe=Salmon!90!Black,
        colback=Salmon!20,
        coltitle=white,
        fonttitle=\large\bfseries,
        title={Optimized Meta-Skill for HLE-MATH (DeepSeek-V4-Flash, Part 2/2)},
        halign title=left,
        fontupper=\footnotesize,
        boxrule=1pt,
        arc=3mm,
        boxsep=2pt,
        left=8pt,
        right=8pt,
        top=4pt,
        bottom=4pt
    ]

\hspace{0em}- Mandatory Self-Consistency and Verification Protocol: Equip every agent with built-in mechanisms for self-consistency checking and fact verification. The prompts must instruct the agent to: (a) upon producing outputs, perform sanity checks (e.g., cross-check with alternative reasoning paths, test boundary conditions, compute bounds if applicable); (b) explicitly state the confidence level for uncertain facts; (c) flag any internal inconsistencies or gaps for downstream resolution. Additionally, each agent's output schema must include a "verification report" that is passed to a dedicated verification agent (defined in Stage 3), making verification a first-class cross-agent contract rather than an optional internal step. This transforms agents from passive executors into active validators, reducing the propagation of hallucinated or erroneous intermediate results. Empirical validation extension: For any derived formula, constant, or specialized fact, agents must perform a mandatory empirical check by testing against at least the smallest feasible known cases (e.g., n=1,2,3 or equivalent boundary conditions) before finalizing. If the test fails, the agent must flag the discrepancy, reconsider the derivation, and optionally consult the MIR for alternative framings. Upgrade: Integrate a Multi-Path Verification requirement: for every critical numeric, logical, or structural output, the agent must produce at least two independent reasoning paths (e.g., theoretical derivation and empirical testing, or symbolic computation and numerical simulation). The verification report must compare both paths, note any discrepancy, and state a reconciled value only when the paths converge. This ensures that observed outputs are robust against single-prompt sensitivity. Extension – Empirical Calibration Protocol: Add a mandatory calibration step for every agent handling numerical, combinatorial, or logical derivation. This protocol requires the agent to: (a) identify the smallest non-trivial instance of the problem; (b) compute the output for that instance using the same reasoning; (c) cross-reference against known ground truth or brute-force enumeration for that instance; (d) compare the small-case result with the full derivation. If a discrepancy is detected, the agent must flag the anomaly, reconsider the derivation, and may trigger a re-derivation loop (up to two iterations). The calibration output must be structured as a JSON report appended to the verification report, enabling cross-verification by downstream agents and providing a low-cost, high-ROI guard against common reasoning errors.

\hspace{0em}- Concession Protocol for Uncertain or Non-Standard Constructs: Each agent's prompts must include a mechanism to detect when a requested term, invariant, or construct is not present in its training knowledge or retrievable from authoritative sources. The agent must be instructed to attempt multiple independent sources (e.g., cross-reference two distinct knowledge bases or reasoning paths). If confidence remains below a high threshold (e.g., no consistent definition can be formulated), the agent must output a formal 'Uncertainty Flag' with status UNKNOWN and confidence 0.0, halting further computation on that sub-task. The pipeline must then bypass the affected sub-task and output a placeholder 'Unknown' answer, preventing fabrication. This protocol transforms hallucination risk into a controlled failure mode, preserving system reliability when encountering genuinely obscure or undefined concepts.

\vspace{6pt}
\hspace{0em}\textbf{3. Workflow \& Orchestration Module (The "How")}

\hspace{0em}Core Objective: Wire the distinct agents from Stage 2 into a functional, executable Multi-Agent System (MAS).

\vspace{3pt}
\hspace{0em}- Architectural Topology: You can design the optimal MAS architecture (e.g., Sequential Pipeline, Router-based, Hierarchical, or Blackboard) based on Stage 1's logical dependencies. For those complex but important sub-tasks, you can design localized different topology design using the instantiated agents. For example, you can define iterative loops for those sub-tasks that need a cycle check to ensure high quality. Or you can call the same agent multiple times to generate diverse outputs and give all of them to the following sub-agents.

\hspace{0em}- Dataflow \& State Management: Define the exact I/O mapping. Specify how the output schema of one agent transforms into the input payload/prompt of downstream agents. Determine how global context (memory/state) is maintained.

\hspace{0em}- Executable Generation: Output the final orchestration logic/code structure that binds the agents, tools, and dataflow into a ready-to-run system.

\hspace{0em}- Robustness Patterns: Verification Gate, Output Sanitization, and Recovery Loops: Incorporate a mandatory verification agent as a final quality gate before output formatting. This agent receives the verification reports from all contributors, cross-checks for consistency, and either approves the final answer or triggers a recovery loop (e.g., retry the failing agent with feedback, spawn a parallel alternative branch, or apply a consensus mechanism across diverse instances). After verification, apply a dedicated output-formatting gate: extract the final answer from the agent's response using a prescribed delimiter (e.g., an isolated token such as \texttt{\textbackslash boxed\{...\}}), verify that the extracted token contains a well-formed, non-empty answer, and set the final system output to exactly that token (with no surrounding prose or formatting). If the delimiter is missing, malformed, or the answer appears incomplete, trigger a retry loop with explicit instructions to produce the answer in the required format.

\hspace{0em}\hspace{2em}- Deterministic Format Enforcement Gate: The formatting gate operates as a rule-based extraction routine: (1) scan the entire output for the last occurrence of the prescribed delimiter; (2) extract the inner content; (3) reconstruct a clean string containing only the delimiter and the extracted content, with no surrounding text; (4) validate that the string matches the required format (e.g., exactly \texttt{\textbackslash boxed\{\textless content\textgreater\}}); (5) if validation fails, invoke a recovery routine that resynthesizes the answer with strict formatting instructions using verified intermediate results. This gate is deterministic and does not depend on the agent's expressive style. It ensures that correct reasoning is never penalized by formatting irregularities. Additionally, the verification gate must include a Cross-Agent Consistency Check that compares verification reports from multiple agents to detect discordant outputs, preventing a single erroneous agent from dominating the final answer. This compound pattern—verification gate, deterministic formatting, structured extraction, and recovery loops—prevents error propagation, mitigates formatting-induced scoring failures, and adds resilience without requiring domain-specific heuristics.

    \end{tcolorbox}
    \caption{Illustration of the optimized Meta-Skill for HLE-MATH (DeepSeek-V4-Flash, Part 2/2).}
    \label{fig:skill_hlemath_2}
\end{figure*}

\begin{figure*}[htbp]
    \centering
    \begin{tcolorbox}[
        enhanced,
        colframe=Salmon!90!Black,
        colback=Salmon!20,
        coltitle=white,
        fonttitle=\large\bfseries,
        title={Optimized Meta-Skill for BrowseComp-Plus (DeepSeek-V4-Flash, Part 1/2)},
        halign title=left,
        fontupper=\footnotesize,
        boxrule=1pt,
        arc=3mm,
        boxsep=2pt,
        left=8pt,
        right=8pt,
        top=4pt,
        bottom=4pt
    ]

\hspace{0em}---

\hspace{0em}name: unified\_meta\_agent\_skill

\hspace{0em}description: "A foundational meta-agent skill for generating Multi-Agent Systems (MAS). It systematically drives the process from conceptual task decomposition to agent engineering, and finally to workflow orchestration."

\hspace{0em}tags:

\hspace{2em}- meta-agent

\hspace{2em}- task-decomposition

\hspace{2em}- agent-engineering

\hspace{2em}- workflow-orchestration

\hspace{0em}inputs:

\hspace{2em}- user\_query

\hspace{0em}---

\vspace{6pt}
\hspace{0em}\textbf{1. Task Decomposition Module (The "What")}

\hspace{0em}Core Objective: Analyze the user query and break it down into a logical blueprint.

\vspace{3pt}
\hspace{0em}- Intent \& Scope Analysis: Understand the macro objective, identify core requirements, and define the boundaries of the task.

\hspace{0em}- Sub-task Breakdown: Decompose the high-level request into a set of discrete, manageable, and logically cohesive sub-tasks.

\hspace{0em}- Constraint Prioritization and Evidence Weighting: For multi-constraint queries, assign each constraint a discriminative weight based on its specificity and independence. Define a clear rule for resolving conflicts when clues point to different answers: prefer the candidate that satisfies the most constraints, with higher weight given to more unique or rare clues. Document which constraints are essential versus complementary to guide downstream evidence scoring.

\hspace{0em}- Logical Dependency Mapping: Identify the business-logic relationships between sub-tasks (e.g., prerequisite, parallel, or iterative). Note: This focuses on logical order, not system dataflow.

\hspace{0em}- Uncertainty and Hypothesis Management: For queries involving multiple implicit or conflict-prone constraints, explicitly structure the decomposition to include parallel retrieval paths for independent clues and a cross-validation node that maintains multiple candidate hypotheses until evidence is sufficient to eliminate alternatives. This prevents premature commitment and confirmation bias.

\hspace{0em}- Constraint Grouping and Bounded Parallelism: When the query contains multiple independent, verifiable constraints, group them into a maximum of four cohesive bundles, each covering at least two related constraints. This prevents over-fragmentation that leads to empty retrieval. For queries with three or more constraints, design a fan-out topology with at most four parallel retrieval sub-tasks, each responsible for a constraint bundle, followed by a single merge sub-task. This balances evidence diversity with retrieval reliability.

\hspace{0em}- Cross-Entity Bridging: When the decomposition involves evidence from two or more distinct entities that must be linked by a specific relationship (e.g., entity A was involved in event B), add a dedicated 'link verification' sub-task that actively searches for the explicit relationship using constraint-aware queries. This sub-task should accept candidate name lists from upstream parallel agents and perform multiple targeted queries to establish the connection. Ensure that the bridging sub-task is assigned to an agent with independent search capability and is placed after the parallel retrieval of entity-specific evidence but before the final consolidation.

\hspace{0em}\textbf{2. Agent Engineering Module (The "Who")}

\hspace{0em}Core Objective: Design specialized sub-agents tailored for the sub-tasks defined in Stage 1.

\hspace{0em}- Role Profiling: Assign a unique identity and specialized role to each sub-agent based on its target sub-task.

\hspace{0em}- Weighted Constraint Satisfaction Protocol with Partial-Evidence Fallback: Every agent that evaluates or synthesizes evidence must be instructed to: (a) assign a weight (e.g., 1–10) to each constraint based on distinctiveness and independence, (b) for each candidate, sum the weights of all SUPPORTED constraints, (c) output the top candidate if its score reaches a clear threshold (e.g., >=70\% of the maximum possible total weight) and no other candidate has a higher score, (d) If no candidate passes the primary threshold, identify the candidate with the highest score that is at least 50\% of the maximum possible total weight. Output that candidate as a "Best guess: [answer] (unverified constraints: [list])" with an explicit confidence note. If no candidate reaches 50\%, output exactly "UNANSWERABLE". (e) Document for each candidate which constraints are supported, missing, or contradicted. All agents (including merge agents) must have independent information-gathering capability to resolve gaps when evidence is incomplete.

\hspace{0em}- Iterative Search Protocol for Retrieval Agents: Any agent responsible for gathering evidence from external sources must follow a structured multi-query search protocol embedded in its role instructions. The protocol mandates at least three distinct attempts: (1) an initial query using the original constraint phrase; (2) if fewer than two relevant results are obtained, relax the least specific constraint and reformulate the query; (3) if results remain insufficient, use the best candidate name from any partial evidence as an anchor and combine it with the remaining constraints to perform a focused search. After three attempts, output a structured uncertainty object (e.g., an evidence summary with explicit gaps) rather than an empty list. This replaces single-shot retrieval and dramatically reduces false negatives.

\hspace{0em}- Verification Autonomy: Any agent tasked with verifying, cross-referencing, or filtering candidates against constraints must be granted independent information-gathering capability (e.g., the ability to perform additional searches) to resolve ambiguous or incomplete evidence. Without this, the agent becomes a reasoning bottleneck that can only operate on pre-fetched data, leading to hallucianted rejections or false confirmations.

\hspace{0em}- Input Context Framing: Specify what contextual information this agent requires from the user or the global task to begin its work.

    \end{tcolorbox}
    \caption{Illustration of the optimized Meta-Skill for BrowseComp-Plus (DeepSeek-V4-Flash, Part 1/2).}
    \label{fig:skill_bcp_1}
\end{figure*}

\begin{figure*}[htbp]
    \centering
    \begin{tcolorbox}[
        enhanced,
        colframe=Salmon!90!Black,
        colback=Salmon!20,
        coltitle=white,
        fonttitle=\large\bfseries,
        title={Optimized Meta-Skill for BrowseComp-Plus (DeepSeek-V4-Flash, Part 2/2)},
        halign title=left,
        fontupper=\footnotesize,
        boxrule=1pt,
        arc=3mm,
        boxsep=2pt,
        left=8pt,
        right=8pt,
        top=4pt,
        bottom=4pt
    ]

\hspace{0em}\textbf{3. Workflow \& Orchestration Module (The "How")}

\hspace{0em}Core Objective: Wire the distinct agents from Stage 2 into a functional, executable Multi-Agent System (MAS).

\vspace{3pt}
\hspace{0em}- Architectural Topology: You can design the optimal MAS architecture (e.g., Sequential Pipeline, Router-based, Hierarchical, or Blackboard) based on Stage 1's logical dependencies. For complex but important sub-tasks, you can design localized different topology designs using the instantiated agents. For example, you can define iterative loops for sub-tasks that need a cycle check to ensure high quality. Or you can call the same agent multiple times to generate diverse outputs and give all of them to the following sub-agents. When the task involves two or more independent constraints that require separate retrieval, enforce a fan-out (parallel) topology with bounded parallelism to protect against single-agent failures, followed by a merge node with cross-validation.

\hspace{0em}- Backtracking and Dynamic Replanning: Design the workflow to detect when a retrieval sub-task returns empty or low-confidence results. Upon detection, trigger a dynamic replanning process: relax constraints (e.g., use broader terms, drop the least discriminative constraint), re-route the sub-task to a more capable agent, or spawn an independent final agent with full search capability that can attempt multiple diverging queries. This prevents catastrophic pipeline collapse and mirrors the pattern of high-performing trajectories where the final agent compensates for upstream gaps.

\hspace{0em}- Merge Node Re-execution Authority: Grant the merge node the authority to independently initiate new retrieval attempts for specific constraints when upstream outputs are incomplete or contradictory. This is achieved by providing the merge agent with access to the search tool and instructing it to re-query for missing evidence using targeted, candidate-anchored queries. This pattern transforms the merge node from a passive aggregator into an active investigator, directly countering cascading empty-output propagation and ensuring that partial evidence does not lead to total failure.

\hspace{0em}- Dataflow \& State Management: Define the exact I/O mapping. Specify how the output schema of one agent transforms into the input payload/prompt of downstream agents. Determine how global context (memory/state) is maintained.

\hspace{0em}- Executable Generation: Output the final orchestration logic/code structure that binds the agents, tools, and dataflow into a ready-to-run system.

    \end{tcolorbox}
    \caption{Illustration of the optimized Meta-Skill for BrowseComp-Plus (DeepSeek-V4-Flash, Part 2/2).}
    \label{fig:skill_bcp_2}
\end{figure*}

\begin{figure*}[htbp]
    \centering
    \begin{tcolorbox}[
        enhanced,
        colframe=Salmon!90!Black,
        colback=Salmon!20,
        coltitle=white,
        fonttitle=\large\bfseries,
        title={Optimized Meta-Skill for VitaBench (DeepSeek-V4-Flash, Part 1/2)},
        halign title=left,
        fontupper=\footnotesize,
        boxrule=1pt,
        arc=3mm,
        boxsep=2pt,
        left=8pt,
        right=8pt,
        top=4pt,
        bottom=4pt
    ]

\hspace{0em}---

\hspace{0em}name: unified\_meta\_agent\_skill

\hspace{0em}description: "A foundational meta-agent skill for generating Multi-Agent Systems (MAS). It systematically drives the process from conceptual task decomposition to agent engineering, and finally to workflow orchestration."

\hspace{0em}tags:

\hspace{2em}- meta-agent

\hspace{2em}- task-decomposition

\hspace{2em}- agent-engineering

\hspace{2em}- workflow-orchestration

\hspace{0em}inputs:

\hspace{2em}- user\_query

\hspace{0em}---

\hspace{0em}\textbf{1. Task Decomposition Module (The "What")}

\hspace{0em}Core Objective: Analyze the user query and break it down into a logical blueprint.

\vspace{3pt}
\hspace{0em}- Intent \& Scope Analysis: Understand the macro objective, identify core requirements, and define the boundaries of the task.

\hspace{0em}- Sub-task Breakdown: Decompose the high-level request into a set of discrete, manageable, and logically cohesive sub-tasks.

\hspace{0em}\hspace{2em}- Granularity Boundary Principle: Any sub-task that inherently spans multiple distinct capability domains—such as information acquisition (search, retrieval) and transactional execution (order, book, purchase)—must be decomposed at the capability boundary into separate sub-tasks. The acquisition sub-task is responsible solely for gathering; the transactional sub-task receives options as an explicit dependency and performs the state-changing action. This prevents a single agent from handling both capacities, which can lead to tool-call failure, and ensures that each sub-task has a clear, focused responsibility. Additionally, for tasks with multiple independent transaction chains (e.g., unrelated purchases or bookings), each chain must start from a clean copy of the original global context, with no cross-chain propagation of outputs unless explicit coordination is required.

\hspace{0em}\hspace{2em}- Acquisition Phase Deep Decomposition: Within any acquisition sub-task that involves selecting a single entity from a set of candidates (e.g., venues, services, products), the acquisition phase must be further decomposed into two distinct nodes: (1) an exploration node that retrieves a diverse set of candidates using broad queries, and (2) an evaluation/selection node that applies multi-criteria scoring against all known constraints before committing to a final choice. This separation prevents premature commitment to a single heuristic and allows systematic comparison.

\hspace{0em}\hspace{2em}- Implicit Constraint Elicitation: Systematically surface constraints that are implied but not explicitly stated in the query (e.g., evaluation rubrics, hidden criteria) and embed them as explicit decision checkpoints in the subtask graph.

\hspace{0em}- Constraint Verification Matrix: Construct a formal matrix that cross-references every known constraint (numeric, categorical, temporal, priority-based) with the sub-tasks it affects. For each constraint, explicitly define its source (user query, implicit rubric, environment), its value, and its precedence over other constraints. This matrix is inherited by all downstream agents as a mandatory context attachment, ensuring that no constraint is forgotten or reinterpreted.

\hspace{0em}- Logical Dependency Mapping: Identify both the business-logical order relationships (prerequisite, parallel, iterative) and the critical data dependencies (e.g., temporal anchors, derived identifiers) that must be transferred between sub-tasks. Map these as a directed graph where edges represent both sequencing and variable handoffs.

\hspace{0em}- Success Criteria: Define clear objective outcomes for each sub-task to ensure evaluability.

\hspace{0em}\textbf{2. Agent Engineering Module (The "Who")}

\hspace{0em}Core Objective: Design specialized sub-agents tailored for the sub-tasks defined in Stage 1.

\hspace{0em}- Role Profiling: Assign a unique identity and specialized role to each sub-agent based on its target sub-task.

\hspace{0em}- Instruction Design: Draft precise system prompts/instructions. Define the agent's specific goals, behavioral boundaries, and output expectations.

\hspace{0em}- Decision-Centric Agent Design: For subtasks involving ambiguous entity selection from candidate sets or conditional branching, design a dedicated "Selector" agent whose single responsibility is to match candidate attributes against all known constraints and output a definitive choice. Critically, all such decision-making agents MUST produce a structured output contract (e.g., a JSON object containing a "decision" key and any extracted numerical or categorical values). Downstream agents receive this contract as their entire contextual payload and use internal reasoning to act upon it, while orchestration code must never parse or inspect free-text output for conditional logic. If the agent cannot produce a valid structured contract, it must request clarification rather than guessing.

\hspace{0em}\hspace{2em}- Candidate Verification Protocol: For any search or retrieval agent that must select an entity from candidate results, the agent's system prompt must include a mandatory verification step. After receiving tool results, the agent must list all candidate entities with their identifiers, evaluate each against all known constraints (from the Constraint Verification Matrix inherited from Stage 1), explicitly state which entity is selected, and provide a rationale that cites the matching constraints. If no candidate fully satisfies all constraints, the agent must request clarification rather than making a best-effort guess. This protocol prevents cascade failures from ambiguous entity selection.

\hspace{0em}\hspace{2em}- Holistic Multi-Criteria Evaluation Mandate: The evaluation step within any selection agent must explicitly score each candidate across all known constraint dimensions (numerical thresholds, categorical matches, temporal boundaries, priority weighting). The agent must document trade-offs and identify the candidate that best satisfies the full set of constraints, not over-weight a single dimension. If multiple candidates equally satisfy all constraints, the agent may select based on a deterministic tiebreaker (e.g., alphabetical ID) but must note the tie in the rationale.

    \end{tcolorbox}
    \caption{Illustration of the optimized Meta-Skill for VitaBench (DeepSeek-V4-Flash, Part 1/2).}
    \label{fig:skill_vita_1}
\end{figure*}

\begin{figure*}[htbp]
    \centering
    \begin{tcolorbox}[
        enhanced,
        colframe=Salmon!90!Black,
        colback=Salmon!20,
        coltitle=white,
        fonttitle=\large\bfseries,
        title={Optimized Meta-Skill for VitaBench (DeepSeek-V4-Flash, Part 2/2)},
        halign title=left,
        fontupper=\footnotesize,
        boxrule=1pt,
        arc=3mm,
        boxsep=2pt,
        left=8pt,
        right=8pt,
        top=4pt,
        bottom=4pt
    ]

\hspace{0em}- Decision Agent Contract for Conditional Logic: For any sub-task that involves a binary or multi-choice conditional decision based on natural-language context (e.g., weather, sentiment, availability text), design the corresponding agent as a specialized Decision Agent whose sole output is a structured decision contract (e.g., boolean with rationale). The orchestration layer must not inspect raw text for decision logic; it passes the contract to downstream routing. This pattern decouples decision intelligence from code logic, making the pipeline resilient to language variation.

\hspace{0em}- Input Context Framing: Specify what contextual information this agent requires from the user or the global task to begin its work.

\vspace{6pt}
\hspace{0em}\textbf{3. Workflow \& Orchestration Module (The "How")}

\hspace{0em}Core Objective: Wire the distinct agents from Stage 2 into a functional, executable Multi-Agent System (MAS).

\vspace{3pt}
\hspace{0em}- Architectural Topology: Design the optimal MAS architecture (e.g., Sequential Pipeline, Router-based, Hierarchical, or Blackboard) based on Stage 1's logical dependencies. For complex but important sub-tasks, you can design localized different topology design using the instantiated agents. For example, you can define iterative loops for those sub-tasks that need a cycle check to ensure high quality, or call the same agent multiple times to generate diverse outputs and provide all of them to the following sub-agents.

\hspace{0em}- Minimum Viable Pipeline Topology: The generated workflow MUST instantiate at least one agent per distinct logical domain (as identified in Stage 1 sub-tasks). The orchestration must include explicit agent invocation steps with tool-call capacities; a monolithic agent that subsumes all domains is forbidden. Validate the orchestration against a structural checklist: presence of multiple agents, at least one tool invocation per domain, and inclusion of retry/fallback mechanisms. This prevents catastrophic pipeline failures due to over-consolidation.

\hspace{0em}- Context Integrity Guarantee: Implement a "Reality Check" gate before any state-changing action (e.g., booking, scheduling) that injects the current global context (e.g., system time, user identity) and explicitly forces verification of temporal or identity parameters against that context, rejecting out-of-range values.

\hspace{0em}- Dataflow \& State Management: Define the exact I/O mapping. Specify how the output schema of one agent transforms into the input payload/prompt of downstream agents. Determine how global context (memory/state) is maintained.

\hspace{0em}\hspace{2em}- Robust Extraction of Environmental Parameters: When the workflow involves conditional branching based on numerical values (e.g., prices, dates, counts) derived from user query, simulation state, or external tool outputs, the orchestration must employ a confidence-based two-stage extraction mechanism. First, attempt targeted pattern matching specific to the parameter type (e.g., structured regular expressions for price or date formats). If the structured extraction fails or returns ambiguous results, delegate the extraction to a dedicated agent with instruction to produce a structured numerical or temporal value from the raw context. The orchestration must never fall back to a brittle default value; it must signal extraction failure if both stages cannot produce a confident value. For temporal references that are expressed as named events (e.g., holidays, festivals) or relative terms (e.g., "tomorrow", "next week"), the orchestration must mandate resolution via a dedicated tool lookup (e.g., holiday date API) rather than heuristic inference from simulation time. This prevents irreversible wrong-branch execution from incorrect parameter parsing.

\hspace{0em}- Executable Generation: Output the final orchestration logic/code structure that binds the agents, tools, and dataflow into a ready-to-run system.

\hspace{0em}\hspace{2em}- Post-Execution Verification Gate: After all sub-task agents have completed, the orchestration must include a verification step that systematically compares each agent's output against a checklist derived from the original task requirements and the Constraint Verification Matrix. If any requirement is found to be unmet, the orchestration must either trigger a targeted re-execution of the responsible sub-task (with additional context from the verification failure) or escalate to a human reviewer. This gate prevents a single early error from cascading into catastrophic pipeline failure and ensures the final aggregated report includes only validated outputs.

\hspace{0em}- Mandatory Final Reporting: Include a dedicated post-execution step (either an additional agent or orchestration logic) that aggregates all agent outputs and formats a comprehensive, user-readable summary. This summary must present all retrieved information (e.g., details of selections, confirmations, status) and transaction results, ensuring no critical information is omitted from the final response.

    \end{tcolorbox}
    \caption{Illustration of the optimized Meta-Skill for VitaBench (DeepSeek-V4-Flash, Part 2/2).}
    \label{fig:skill_vita_2}
\end{figure*}

\begin{table*}[t]
\centering
\small
\begin{tabular}{@{}>{\raggedright\arraybackslash}m{0.14\linewidth}>{\centering\arraybackslash}m{0.50\linewidth}>{\raggedright\arraybackslash}m{0.3\linewidth}@{}}
\toprule
\textbf{Methos} & \textbf{MAS Code (Input $\rightarrow$ Output)} & \textbf{Structure Description} \\
\midrule
\multicolumn{3}{@{}p{\linewidth}@{}}{\textbf{Input Query:} Using these hints which are all correct as of December 31 2023, Identify the athlete  1. During their career, the person played in the same team with a player whose father scored 33 goals for that same team.  2. The person played 11 games in a major tournament with a goal average of 0.55 3. The player won a pro league title between 2010 and 2020  4. The player was featured on a single in the same year they scored 6 goals for the national team 5. They also represented their nation only once at a certain major international sporting event.}\\
\midrule
\textbf{EvoAgent} & \diagEvoBCP & Expert planners retrieve in parallel, but \rneg{they search for the same query without considering decomposition and global constraints}. \bad \\
\textbf{AOrchestra} & \diagAOrchBCP & MainAgent + delegate/sub-agent can iterate search, but \rneg{it is similar to the vanilla multi-turn search and does not introduce agentic elements}. \bad \\
\textbf{AFlow} & \diagAFlowBCP & Search/SC/verify loop improves robustness, but \rneg{the multiple searches are identical and verification remains answer-level, not entity-link-level across multiple searches}. \bad \\
\textbf{MAS$^2$} & - & \rneg{Failed, it cannot generate executable MAS code}. \bad \\
\textbf{MAS-Orchestra} & \diagOrchBCP & Multi-branch debate + reflexion explores alternatives, \rneg{although the direction of the analysis is correct and detailed, it is limited by the fact that only one search has been designed, resulting in insufficient information}. \bad \\
\midrule
\textbf{Skill-MAS-init} & \diagInitBCP & Linear hint filtering keeps only one evidence path, so \rneg{early mismatch is hard to recover and final athlete selection drifts}. \bad \\
\midrule
\textbf{Skill-MAS-optimized} & \diagOptBCP & \gpos{\textbf{Structural advantage:} \texttt{parse\_and\_plan} splits into five clue-specific retrievers, then \texttt{link\_verification} enforces cross-clue consistency before \texttt{merge\_and\_decide}.} \good \\
\bottomrule
\end{tabular}
\caption{MAS structure comparison on BrowseComp-Plus (DeepSeek-V4-Flash).}
\label{tab:mas-compare-bcp}
\end{table*}

\begin{table*}[t]
\centering
\small
\begin{tabular}{@{}>{\raggedright\arraybackslash}m{0.14\linewidth}>{\centering\arraybackslash}m{0.50\linewidth}>{\raggedright\arraybackslash}m{0.3\linewidth}@{}}
\toprule
\textbf{Method} & \textbf{MAS Code (Input $\rightarrow$ Output)} & \textbf{Structure Description} \\
\midrule
\multicolumn{3}{@{}p{\linewidth}@{}}{\textbf{Input Query:} Complete You have a day off today. You did a thorough cleaning at home and sweated all over. You are really too lazy to cook. You plan to order some light food for takeout. If you want to have a light and refreshing taste, it should include meat, vegetables and fruits. But you should avoid pork and beef, and also stay away from high-purine and caffeine-containing foods. It should be delivered around 13:00. You were recently recommended a Nobel Prize-winning novel, but you always can't get into it when reading at home. After lunch, you want to see if there are any quiet books nearby. It should be more atmospheric. When reading, you want to have some tea to refresh yourself. You want to see what tea packages are available in the book bar. You don't like black tea and still want an independent reading space. If there is a suitable one, you can buy a coupon first. I'm going to Yuncheng on a business trip by train tomorrow afternoon. You want to buy a train around 3 o 'clock. If the first-class seat doesn't cost more than 200 yuan, then buy a first-class seat. Otherwise, it's better to go for a second-class seat.}\\
\midrule
\textbf{EvoAgent} & \diagEvoVita & Multi-expert tool loops collect options, but \rneg{handoffs are not explicitly staged as explore/evaluate pairs, so purchase and booking states are inconsistent across experts}. \bad \\
\textbf{AOrchestra} & \diagAOrchVita & Main/delegate + env bridge is flexible, yet \rneg{it lacks constraints checking like budget for each sub-task branch}. \bad \\
\textbf{AFlow} & \diagAFlowVita & Two-stage \texttt{Custom} generate$\rightarrow$verify, but \rneg{single linear chain has no per-subtask explore/evaluate/order branches and is easy to miss some tasks}. \bad \\
\textbf{MAS$^2$} & \diagMASSVita & Linear workflow, \rneg{tool execution is deferred to a single post-reasoning bridge instead of staged per sub-task transactions}. \bad \\
\textbf{MAS-Orchestra} & \diagOrchVita & One SCAgent, \rneg{the multiple CoT calls are for the same query without problem decomposition and per-subtask branches}. \bad \\
\midrule
\textbf{Skill-MAS-init} & \diagInitVita & Compressed 4-step flow (\texttt{order\_meal -> find\_book\_cafe -> purchase\_voucher -> book\_train\_ticket}) mixes retrieval, screening, and transaction in a linear chain, so \rneg{it is sensitive to error propagation, text contamination, and constraints requirements}. \bad \\
\midrule
\textbf{Skill-MAS-optimized} & \diagOptVita & \gpos{\textbf{Structural advantage:} \texttt{constraint\_extraction} dispatches three explicit explore$\rightarrow$evaluate$\rightarrow$order branches (meal/bookbar/train), and \texttt{final\_summary} aggregates only after all branch decisions complete.} \good \\
\bottomrule
\end{tabular}
\caption{MAS structure comparison on VitaBench (DeepSeek-V4-Flash).}
\label{tab:mas-compare-vita}
\end{table*}

\section{Prompt Details}
\label{appendix:prompt details}

We elaborate on prompts used in Skill-MAS from Figure~\ref{fig:llm_judge_drb} to Figure~\ref{fig:skill_mas_optimization}. These instructions cover LLM-as-judge evaluation and the entire framework pipeline, including Skill-MAS building, selective reflection, and skill optimization.

\begin{figure*}[htbp]
    \centering
    \begin{tcolorbox}[
        enhanced,
        colframe=Salmon!90!Black,
        colback=Salmon!20,
        coltitle=white,
        fonttitle=\large\bfseries,
        title={LLM-as-a-judge Prompts (DeepResearchBench)},
        halign title=left,
        fontupper=\footnotesize,
        boxrule=1pt,
        arc=3mm,
        boxsep=2pt,
        left=8pt,
        right=8pt,
        top=4pt,
        bottom=4pt
    ]

\hspace{0em}\textbf{System Prompt:}
You are a strict, meticulous, and objective research article evaluation expert. You excel at using specific assessment criteria to deeply compare two articles on the same task, providing precise scores and clear justifications.

\hspace{0em}\textbf{User Prompt:}

\textbf{Task Background:}
There is a deep research task, and you need to evaluate two research articles written for this task. We will assess the articles across four dimensions: Comprehensiveness, Insight, Instruction Following, and Readability. The content is as follows:

\hspace{4em}\texttt{<task>"\{task\_prompt\}"</task>}

\hspace{2em}\textbf{Articles to Evaluate:}

\hspace{4em}\texttt{<article\_1>"\{article\_1\}"</article\_1>}

\hspace{4em}\texttt{<article\_2>"\{article\_2\}"</article\_2>}

\hspace{2em}\textbf{Evaluation Criteria:}

\hspace{4em}Now, you need to evaluate and compare these two articles based on the following evaluation criteria list, providing comparative analysis and scoring each on a scale of 0-10. Each criterion includes an explanation, please understand carefully.
\hspace{0em}\texttt{<criteria\_list>\{criteria\_list\}</criteria\_list>}

\hspace{2em}\textbf{Instruction:}

\hspace{4em}\textbf{Your Task}

\hspace{4em}Please strictly evaluate and compare \texttt{<article\_1>} and \texttt{<article\_2>} based on each criterion in the \texttt{<criteria\_list>}. You need to:

\hspace{6em}1. Analyze Each Criterion: Consider how each article fulfills the requirements of each criterion.

\hspace{6em}2. Comparative Evaluation: Analyze how the two articles perform on each criterion, referencing the content and criterion explanation.

\hspace{6em}3. Score Separately: Based on your comparative analysis, score each article on each criterion (0-10 points).

\hspace{4em}\textbf{Scoring Rules}

\hspace{6em}For each criterion, score both articles on a scale of 0-10 (continuous values). The score should reflect the quality of performance on that criterion:

\hspace{6em}- 0-2 points: Very poor performance. Almost completely fails to meet the criterion requirements.

\hspace{6em}- 2-4 points: Poor performance. Minimally meets the criterion requirements with significant deficiencies.

\hspace{6em}- 4-6 points: Average performance. Basically meets the criterion requirements, neither good nor bad.

\hspace{6em}- 6-8 points: Good performance. Largely meets the criterion requirements with notable strengths.

\hspace{6em}- 8-10 points: Excellent/outstanding performance. Fully meets or exceeds the criterion requirements.

\hspace{4em}\textbf{Output Format Requirements}

\hspace{6em}Please strictly follow the \texttt{<output\_format>} below for each criterion evaluation. Do not include any other unrelated content, introduction, or summary. Start with "Standard 1" and proceed sequentially through all criteria.

\hspace{4em}\texttt{<output\_format>}

\hspace{4em}\texttt{\{}
\hspace{2em}\texttt{"comprehensiveness": [}

\hspace{8em}\texttt{\{}

\hspace{10em}\texttt{"criterion": [Text content of the first comprehensiveness evaluation criterion],}

\hspace{10em}\texttt{"analysis": [Comparative analysis],}

\hspace{10em}\texttt{"article\_1\_score": [Continuous score 0-10],}

\hspace{10em}\texttt{"article\_2\_score": [Continuous score 0-10]}

\hspace{8em}\texttt{\},}

\hspace{8em}\texttt{\{}

\hspace{10em}\texttt{"criterion": [Text content of the second comprehensiveness evaluation criterion],}

\hspace{10em}\texttt{"analysis": [Comparative analysis],}

\hspace{10em}\texttt{"article\_1\_score": [Continuous score 0-10],}

\hspace{10em}\texttt{"article\_2\_score": [Continuous score 0-10]}

\hspace{8em}\texttt{\},}

\hspace{8em}\texttt{...}

\hspace{6em}\texttt{],}

\hspace{6em}\texttt{"insight": [}

\hspace{8em}\texttt{\{}

\hspace{10em}\texttt{"criterion": [Text content of the first insight evaluation criterion],}

\hspace{10em}\texttt{"analysis": [Comparative analysis],}

\hspace{10em}\texttt{"article\_1\_score": [Continuous score 0-10],}

\hspace{10em}\texttt{"article\_2\_score": [Continuous score 0-10]}

\hspace{8em}\texttt{\},}

\hspace{8em}\texttt{...}

\hspace{6em}\texttt{],}

\hspace{6em}\texttt{...}
\hspace{2em}\texttt{\}}
\hspace{4em}\texttt{</output\_format>}

\hspace{2em}Now, please evaluate the two articles based on the research task and criteria, providing detailed comparative analysis and scores according to the requirements above. Ensure your output follows the specified output format and that the JSON format is parsable, with all characters that might cause JSON parsing errors properly escaped.

    \end{tcolorbox}
    \caption{LLM-as-a-judge prompts used in DeepResearchBench.}
    \label{fig:llm_judge_drb}
\end{figure*}

\begin{figure*}[htbp]
    \centering
    \begin{tcolorbox}[
        enhanced,
        colframe=Salmon!90!Black,
        colback=Salmon!20,
        coltitle=white,
        fonttitle=\large\bfseries,
        title={LLM-as-a-judge Prompts (VitaBench)},
        halign title=left,
        fontupper=\footnotesize,
        boxrule=1pt,
        arc=3mm,
        boxsep=2pt,
        left=8pt,
        right=8pt,
        top=4pt,
        bottom=4pt
    ]

\hspace{0em}\textbf{System Prompt:}
System Information
\texttt{\{env\_info\}}
User Complete Instruction
\texttt{\{user\_instruction\}}

\textbf{Background}
- This is a conversation scenario between a user and an assistant, where the assistant can call tools to retrieve information and complete operations. Tool return results will start with "tool"

\hspace{4em}- You need to evaluate whether the user instruction has been completed. The user's complete instruction has been broken down into several scoring points (rubrics), and you only need to judge whether each scoring point is satisfied

\hspace{4em}- Due to the large number of conversation turns, we use sliding window evaluation, where each window shows 10 conversation turns with 2 overlapping turns between windows, but rubric status will be preserved across windows

\hspace{4em}- You are evaluating window \texttt{\{window\_idx\}} (out of \texttt{\{total\_windows\}} windows total)

\hspace{4em}- \texttt{<window\_content>} contains the conversation content for the current window

\hspace{4em}- \texttt{<current\_rubrics>} contains the current status of all scoring points (true means satisfied, false means not satisfied, all scoring points have an initial status of false)

\hspace{2em}\textbf{Task}
- Update the scoring point rubric status based on the conversation content in the current window

\hspace{4em}- You can update the status from false to true, if and only if the assistant completed the goal in this window

\hspace{4em}- You can also update true back to false, if and only if the assistant overturned a previous correct conclusion in this window (but note: if the user's needs themselves changed, such as actively canceling after placing an order, the original evaluation of the ordering requirement should not be overturned)

\hspace{4em}- You can refer to the "User Complete Instruction" to understand the progress of the current conversation window and avoid unnecessary modifications

\hspace{2em}\textbf{Important Notes}

\hspace{4em}- Important: All evaluations are based on whether the assistant's responses and tool call requests complete the goals in the rubrics. User expressions in the conversation are only considered as guidance for the assistant and do not directly affect evaluation standards. Everything is based on the rubric fields!

\hspace{4em}- Important: Query tool return results are only visible to the assistant and do not represent content recommended by the assistant to users, so they do not directly affect evaluation results either. Everything must be based on the assistant's responses to users after obtaining information! Also note that the Assistant cannot fabricate Tool return results!

\hspace{4em}- Important: For order-related rubrics (involving order details that must generate orders), you must confirm whether the assistant actually completed the ordering operation. The assistant may mistakenly believe they completed the ordering operation when in fact the tool call failed; or situations where the user states they "can place the order themselves," etc., should all be considered as not meeting the requirements

\hspace{4em}- For rubrics involving order details such as product quantity or delivery time, the original rubric requirements must be strictly met (no deviation in product quantities, delivery must not be later than the expected time). User compromise behavior does not affect evaluation results (for example, when a user states "fewer items is okay", "I have no objections to the order content" or "later delivery is fine" etc.). These situations should still be considered as not meeting the requirements

\hspace{4em}- For rubrics involving text content matching of addresses or order notes, apply the functional equivalence principle: as long as the actual content can achieve the same function (such as roughly locating the delivery location or conveying the customer's main needs), it is considered to meet the requirements even if the expression is not completely consistent or lacks some details

\hspace{4em}- If the current window does not involve a certain rubric, keep its original status unchanged; if the current window involves a rubric but cannot fully determine it, key information can be recorded in the justification for future judgment

\hspace{4em}- In the justification, record key information related to the current rubric and its corresponding round [x] in an appending manner. If status modifications occur, record the reason using concise language. If the status is not modified, restate the previous reason

\hspace{2em}Format Requirements

\hspace{4em}- Your response should be a JSON object containing the following fields:

\hspace{6em}- \texttt{rubric\_idx}: Unique identifier for the rubric

\hspace{6em}- \texttt{rubric}: Restatement of the rubric

\hspace{6em}- \texttt{justification}: Explanation of status changes

\hspace{6em}- \texttt{meetExpectation}: Updated status (true or false)

\hspace{2em}Example Input Structure:

\hspace{4em}\texttt{<window\_content>xxx</window\_content>}

\hspace{4em}\texttt{<current\_rubrics>xxx</current\_rubrics>}

\hspace{2em}Example Response Structure:

\hspace{4em}\texttt{[}
\hspace{0em}\texttt{\{}

\hspace{8em}\texttt{"rubric\_idx": "rubric\_0",}

\hspace{8em}\texttt{"rubric": "<restate the rubric>",}

\hspace{8em}\texttt{"justification": "<brief explanation of status change, recorded in appended format>",}

\hspace{8em}\texttt{"meetExpectation": <true or false>}

\hspace{6em}\texttt{\},}
\hspace{0em}\texttt{...}
\hspace{0em}\texttt{]}

\hspace{0em}\textbf{User Prompt:}
\hspace{2em}\texttt{<window\_content>}
\texttt{\{window\_content\}}
\hspace{0em}\texttt{</window\_content>}

\hspace{2em}\texttt{<current\_rubrics>}
\texttt{\{current\_rubrics\_str\}}
\hspace{0em}\texttt{</current\_rubrics>}

    \end{tcolorbox}
    \caption{LLM-as-a-judge prompts used in VitaBench.}
    \label{fig:llm_judge_vita}
\end{figure*}

\begin{figure*}[htbp]
    \centering
    \begin{tcolorbox}[
        enhanced,
        colframe=Salmon!90!Black,
        colback=Salmon!20,
        coltitle=white,
        fonttitle=\large\bfseries,
        title={Skill-MAS Build Contract (Part 1/2)},
        halign title=left,
        fontupper=\footnotesize,
        boxrule=1pt,
        arc=3mm,
        boxsep=2pt,
        left=8pt,
        right=8pt,
        top=4pt,
        bottom=4pt
    ]

\hspace{0em}\textbf{[GENERAL\_CONSTRAINTS]}

\hspace{0em}- Single JSON object only (no markdown fences, no surrounding prose).

\hspace{0em}- The runner parses the first top-level \texttt{\{...\}} object from your reply.

\hspace{0em}- Key names must match [OUTPUT\_JSON\_SCHEMA]; prior-stage outputs are embedded verbatim into later [STAGE\_INPUT\_JSON] blobs.

\hspace{0em}- Fields such as \texttt{dataset\_final\_requirements} encode benchmark grading contracts, not optional narrative hints.

\hspace{0em}- Cross-Stage Execution Safety: Stage 2 \texttt{role\_instruction} / \texttt{user\_prompt} strings and Stage 3 \texttt{mas\_code} are ultimately embedded into executable Python. Any content that breaks Python string literals or syntax (unescaped quotes, unmatched parentheses, duplicate import blocks, invalid dict keys) will cause total workflow failure at runtime. Design all text so it survives embedding as Python source.

\hspace{0em}\rule{\linewidth}{0.4pt}

\hspace{0em}\textbf{[MAS\_BUILD\_CONTRACT — Stage 1]}

\hspace{0em}\textbf{[STAGE\_SPECIFIC\_CONSTRAINTS]}

\hspace{0em}Stage 1: Task Decomposition — Engineering Constraints.

\hspace{0em}- Node Count Limits: the number of decomposed sub-tasks must be kept within a reasonable range (typically 3-6 for most tasks), because over-fragmentation leads to context window explosion and information dilution, while under-fragmentation defeats the purpose of a MAS.

\hspace{0em}- Strict Node Naming Convention: all sub-task IDs must be globally unique and formatted strictly in lowercase snake\_case (e.g., data\_analyzer). This ID serves as the absolute identifier across all stages, and modifying the spelling or casing in later stages is strictly forbidden.

\hspace{0em}- Decomposition Principle — Describe WHAT, Not HOW: you are defining the structural blueprint of the multi-agent system, NOT solving the user's problem. Each sub-task should describe a capability or processing stage (e.g., 'retrieve\_evidence', 'analyze\_constraints', 'synthesize\_answer'), not attempt to answer the question or provide solution content. The actual solving happens later when the generated agents execute at runtime.

\hspace{0em}- Topology Awareness: consider whether the task benefits from a simple sequential chain, or needs a fan-out/fan-in pattern (multiple independent analyses merged at the end), or an iterative loop (e.g., retrieve → analyze → refine → retrieve again). Express this via the dependency structure. For tasks requiring trial-and-error (e.g., open-domain search), design at least one iterative or fallback-capable node rather than a single one-shot attempt.

\hspace{0em}- Information Preservation: ensure each sub-task's description is self-contained enough that a downstream agent can understand the context without needing the full original query. Avoid designs where intermediate outputs become too thin (e.g., a 'filter' node that outputs only null/empty — design instead for nodes that always produce structured, actionable output even when uncertain).

\hspace{0em}\rule{\linewidth}{0.4pt}

\hspace{0em}\textbf{[MAS\_BUILD\_CONTRACT — Stage 2]}

\hspace{0em}\textbf{[STAGE\_SPECIFIC\_CONSTRAINTS]}

\hspace{0em}Stage 2: Agent Engineering — Engineering Constraints.

\hspace{0em}- Absolute 1-to-1 Node Mapping: every sub-agent generated in this stage must strictly correspond to a sub-task defined in Stage 1; the role\_name must perfectly match the node ID from Stage 1. Do not omit any nodes, and do not hallucinate new agents that were not planned.

\hspace{0em}- Strict Boundary Enforcement: agent instructions and user prompts must NOT contain any routing or scheduling logic. Agents should never be instructed to "pass the result to the reviewer agent." Agents must only focus on completing their specific sub-task. The system framework will handle the data routing.

\hspace{0em}- You Are Engineering Prompts, Not Solving Problems: your output is a set of agent specifications (role\_instruction + user\_prompt) that will be used to call an LLM at runtime. Do NOT attempt to answer the user's question yourself. Your role\_instruction defines WHAT the agent should do and its behavioral boundaries. The user\_prompt must be short (see Runtime Task Context below); the full task is always injected at runtime in \texttt{GLOBAL TASK CONTEXT}, so you do not need to duplicate the entire problem inside \texttt{user\_prompt}.

\hspace{0em}- Runtime Task Context (Critical): at execution time, every sub-agent prompt is assembled with a \texttt{GLOBAL TASK CONTEXT} section that already contains the full original task\_text. Therefore your \texttt{user\_prompt} MUST stay short and Python-embeddable (plain ASCII where possible): instruct the agent to read \texttt{GLOBAL TASK CONTEXT} for the full problem, citations, LaTeX, and output format — do NOT paste the entire task, code fences, or long LaTeX into \texttt{user\_prompt} because Stage 3 copies these strings into \texttt{SubAgentRequest(...)} inside generated Python and unescaped quotes or triple-quote substrings will cause SyntaxError and zero score.

\hspace{0em}- Avoid Empty/Null Output Designs: do NOT design agents whose natural output could be empty, null, or a simple 'not found' message. For example, instead of a 'filter\_candidates' agent that might output null when no candidates match, design an 'analyze\_and\_rank' agent that always produces a structured ranking or analysis. Every agent should produce actionable, substantive output.

\hspace{0em}- Execution Mode Reasoning: choose tool\_context.execution\_mode based on the actual capabilities needed. For tasks requiring information retrieval, use 'multi\_turn\_search' so the agent can perform multi-round searches. For pure reasoning/analysis, use 'llm\_only'. Do not default all agents to 'llm\_only' — the system's effectiveness depends on correctly routing tool-capable agents to the right execution mode.

\hspace{0em}- Upstream-to-Retrieval Bridge (Critical for multi\_turn\_search agents): if a sub-agent with execution\_mode='multi\_turn\_search' depends on an upstream agent, the upstream agent's output must contain actionable retrieval guidance — not just a re-formatted restatement of the original query constraints.

    \end{tcolorbox}
    \caption{MAS build contract used in the three-stage Skill-MAS construction pipeline (Part 1/2).}
    \label{fig:skill_mas_build_1}
\end{figure*}

\begin{figure*}[htbp]
    \centering
    \begin{tcolorbox}[
        enhanced,
        colframe=Salmon!90!Black,
        colback=Salmon!20,
        coltitle=white,
        fonttitle=\large\bfseries,
        title={Skill-MAS Build Contract (Part 2/2)},
        halign title=left,
        fontupper=\footnotesize,
        boxrule=1pt,
        arc=3mm,
        boxsep=2pt,
        left=8pt,
        right=8pt,
        top=4pt,
        bottom=4pt
    ]

The multi\_turn\_search agent uses a BM25 (lexical keyword) retrieval engine under the hood; its planner receives the upstream output as 'Delegation / planning context'. If the upstream output only contains vague abstractions like 'amphibian' or 'early 1990s', the BM25 planner cannot generate precise search queries. Instead, the upstream agent should produce: (a) concrete candidate entity names, specific search keywords, or named entities that BM25 can match directly; (b) a 'suggested\_search\_queries' list of 3-5 distinct BM25-ready queries; (c) any disambiguation clues (e.g., 'the amphibian reference likely means frog/newt/salamander in a company name'). If the upstream agent is a pure constraint extractor and cannot supply such concrete terms, then DO NOT place it before a multi\_turn\_search agent — instead, make the multi\_turn\_search agent the FIRST node in the pipeline so it receives the full original query directly and can derive its own search strategy.

\hspace{0em}\rule{\linewidth}{0.4pt}

\hspace{0em}\textbf{[MAS\_BUILD\_CONTRACT — Stage 3]}

\hspace{0em}\textbf{[STAGE\_SPECIFIC\_CONSTRAINTS]}

\hspace{0em}Stage 3: Workflow Orchestration \& Code Generation — Engineering Constraints.

\hspace{0em}- Template Immutability: you must strictly build upon the provided MASWorkflowTemplate; do not modify the function signatures (arguments or return types) of \texttt{\_\_init\_\_}, \texttt{forward\_async}, or \texttt{build\_workflow\_subagent\_prompt}; only inject code between the designated \texttt{<START>} and \texttt{<END>} markers.

\hspace{0em}- No Third-Party Orchestration Frameworks: try not to import other libraries as much as possible; if you truly need to import external libraries, do not import or attempt to use external MAS frameworks like langgraph, autogen, or crewai; use only Python standard libraries.

\hspace{0em}- No Duplicate Imports: do NOT emit \texttt{from \_\_future\_\_ import annotations}, \texttt{from typing ...}, or \texttt{from template.sub\_agent ...} — the runner prepends a fixed import block. Emit only the workflow class body. Duplicate imports cause invalid Python and immediate failure.

\hspace{0em}- Syntactically Valid Python: before returning \texttt{mas\_code}, mentally verify matching parentheses/brackets/braces, valid string literals, and valid dict keys in \texttt{upstream\_outputs} (use simple snake\_case string keys only, e.g. \texttt{"upstream\_a"} — never use task text fragments or LaTeX as dict keys).

\hspace{0em}- String Literal Safety for SubAgentRequest fields: when embedding \texttt{role\_instruction} and \texttt{user\_prompt} into Python, prefer triple-single-quoted raw strings \texttt{r'''...'''} or \texttt{'''...'''} and ensure the chosen delimiter never appears unescaped inside the text. If you use double-quoted strings, every internal quote must be escaped. Never embed triple-double-quotes inside triple-double-quoted strings.

\hspace{0em}- Safe State Dictionary Access: when populating the upstream\_outputs dictionary in forward\_async, the values must only reference keys in the state dictionary that have ALREADY been executed and populated; do not attempt to access a node's output before it has been awaited, and do not hallucinate variable names.

\hspace{0em}- Force Sequential Await (No Complex Concurrency): even if the DAG topology contains logically parallel branches, you must generate the code using sequential await calls; do not use asyncio.gather(); the DAG dataflow is perfectly maintained by passing the correct variables via upstream\_outputs without needing concurrent execution logic.

\hspace{0em}- Mandatory Final State Assignment (Critical — Single-Source Rule): forward\_async must contain exactly ONE assignment to \texttt{state["final\_output"]}, and it must be of the form \texttt{state["final\_output"] = state["out\_<FINAL\_AGENT\_ROLE\_NAME>"]} — a direct reference to the output of the last/topological-sink sub-agent stored in the state dict. This rule has three strict sub-rules: (1) No hardcoded string literals: do NOT assign a fixed string like \texttt{"UNANSWERABLE"} or \texttt{"EVIDENCE\_INSUFFICIENT"} to \texttt{state["final\_output"]}. The final answer must always come from an agent execution, never from a hardcoded fallback string. If retrieval or analysis fails, the final agent itself must still produce its best-effort answer — a hardcoded UNANSWERABLE string completely bypasses the agent and guarantees a zero score. (2) No f-string concatenation or intermediate variable: do NOT use f-string summary patterns or \texttt{final\_summary = ...; state["final\_output"] = final\_summary}. The final\_output must be exactly \texttt{state["out\_<ROLE>"]} so the runtime can reliably extract the agent's direct output without parsing wrapper text, JSON envelopes, or concatenated summaries. (3) No duplicate assignments: do NOT write two \texttt{state["final\_output"] = ...} lines. Write exactly one assignment. Additionally, assign total token usage to \texttt{state["usage\_final"]} (typically \texttt{state["usage\_<FINAL\_AGENT\_ROLE\_NAME>"]}). The system relies on these keys to evaluate the success of the MAS.

\hspace{0em}- Strict Bounding for Loops and Ensembles (Crucial): if you design advanced topologies like iterative correction loops (e.g., while not approved:) or ensemble generation (calling the same agent multiple times), or other advanced topology, you MUST hardcode a strict upper bound limit. Any loop must be explicitly capped at a maximum of 3 iterations (e.g., using a counter if attempts >= 3: break or for \_ in range(3):). Infinite loops are strictly prohibited.

\hspace{0em}- Safe State Dictionary Access and Loop Variable Handling: when populating upstream\_outputs or updating variables within a loop, values must only reference keys in the state dictionary that have ALREADY been executed; inside loops, ensure you correctly overwrite or append to state keys (e.g., \texttt{state["current\_code\_draft"]}) so the next iteration receives the updated data, and this requirement is the same for ensemble topology or other advanced topologies.

\hspace{0em}- Output Fallback Logic: if a retrieval or tool-based agent produces an empty, null, or otherwise unusable output, the downstream agent should NOT propagate that null downstream. Instead, design the code to detect empty upstream outputs and either (a) re-execute the upstream agent with a modified prompt, or (b) provide a default fallback value with a clear marker. This prevents cascading null/empty failures through the pipeline.

\hspace{0em}- Final Agent Output Guarantee: the last agent in the workflow MUST produce the complete, final answer in the exact format required by the benchmark (see dataset\_final\_requirements in Stage 1). Do NOT let the final agent produce a summary, meta-commentary, or anything other than the direct answer. If the upstream data is insufficient, the final agent should still attempt to produce its best answer based on available information rather than outputting null or empty.

    \end{tcolorbox}
    \caption{MAS build contract used in the three-stage Skill-MAS construction pipeline (Part 2/2).}
    \label{fig:skill_mas_build_2}
\end{figure*}

\begin{figure*}[htbp]
    \centering
    \begin{tcolorbox}[
        enhanced,
        colframe=Salmon!90!Black,
        colback=Salmon!20,
        coltitle=white,
        fonttitle=\large\bfseries,
        title={Skill-MAS Evolution (Within-Task reflection)},
        halign title=left,
        fontupper=\footnotesize,
        boxrule=1pt,
        arc=3mm,
        boxsep=2pt,
        left=8pt,
        right=8pt,
        top=4pt,
        bottom=4pt
    ]

\hspace{0em}\textbf{System Prompt}
You are the diagnosis agent for Skill\_MAS Step 2 (Trajectory Reflection Synthesis). Your task is to analyze trajectories for ONE task at a time, produce that sample's contrastive diagnosis, and output structured JSON (including narrative\_summary for a later cross-sample call). This call does not include other tasks — synthesize only within the provided task\_id.

\hspace{0em}\textbf{User Prompt}
[Step2: Contrastive Trajectory Analysis — Phase 1 (Intra-sample only)]

\hspace{2em}=== SELECTION CONTEXT (this call) ===

\hspace{4em}task\_id: \{task\_id\}
\hspace{0em}rollouts for this task: \{num\_rollouts\}
\hspace{0em}Input statistics:
\hspace{0em}- Character count: \{input\_char\_count\}
\hspace{0em}- Estimated tokens: \textasciitilde\{estimated\_input\_tokens\}
\hspace{0em}Trajectory indices: the JSON field trajectories[] is sorted by score ascending. Use 0-based indices into trajectories[] for high\_performing\_indices / low\_performing\_indices.

\hspace{4em}Selection rationale: This sample was selected for HIGH PRIORITY due to:
\hspace{0em}1. High cross-trajectory volatility: Large score variance across rollouts indicates unstable/inconsistent policy behavior
\hspace{0em}2. High intrinsic difficulty: Low average scores suggest systematic capability gaps

\hspace{4em}Your mission: Diagnose WHY this sample is volatile and/or difficult, and propose an actionable fix for this task.

\hspace{2em}=== INPUT DATA ===

\hspace{4em}Below are ALL trajectories (this task, all k rollouts):
\hspace{0em}\{trajectories\_payload\}

\hspace{2em}=== ANALYSIS FRAMEWORK ===

\hspace{4em}LEVEL 1: INTRA-SAMPLE CONTRASTIVE ANALYSIS (Per-Sample Deep Dive)

\hspace{4em}For THIS sample, perform contrastive analysis by comparing its trajectories:

\hspace{4em}Step 1.1: Quantify the divergence
\hspace{0em}- Report score distribution: min, max, mean, std, gap (max-min)
\hspace{0em}- Identify which trajectories are "high-performing" vs "low-performing" (use the median score as the threshold)
\hspace{0em}- Calculate volatility metrics if applicable

\hspace{4em}Step 1.2: Contrastive diagnosis (CRITICAL - be specific!)
\hspace{0em}Answer these questions by comparing concrete trajectory behaviors:

\hspace{6em}\textbf{A. Divergence points}: Where do high-score and low-score trajectories START to diverge?
\hspace{0em}- Identify the specific step/phase/decision where paths split
\hspace{0em}- What different choices/actions/strategies do they take?
\hspace{0em}- Is the divergence due to: exploration randomness, reasoning errors, constraint violations, or something else?
\hspace{0em}\textbf{B. Success factors}: Why do high-score trajectories succeed?
\hspace{0em}- What specific behaviors/strategies/patterns lead to success?
\hspace{0em}- Are there critical decisions that high-score trajectories get right?
\hspace{0em}- Is success due to: correct reasoning, better exploration, constraint adherence, or luck?
\hspace{0em}\textbf{C. Failure modes}: Why do low-score trajectories fail?
\hspace{0em}- What specific errors/mistakes/violations occur?
\hspace{0em}- Are failures due to: wrong reasoning, premature termination, constraint violations, inefficient search, or capability gaps?
\hspace{0em}- Are failures recoverable (wrong path but could backtrack) or fundamental (missing capability)?
\hspace{0em}\textbf{D. Volatility root cause}: If this sample has high uncertainty, what causes it?
\hspace{0em}- Is the task inherently ambiguous (multiple valid interpretations)?
\hspace{0em}- Is the policy's decision-making unstable at critical junctions?
\hspace{0em}- Are there stochastic elements (exploration, sampling) that amplify variance?
\hspace{0em}\textbf{E. Difficulty root cause}: If this sample has low average score, what makes it hard?
\hspace{0em}- Does it require capabilities the policy lacks (complex reasoning, long-term planning, domain knowledge)?
\hspace{0em}- Does it have tight constraints that are easy to violate?
\hspace{0em}- Is it a boundary case that exposes edge-case weaknesses?

\hspace{4em}Step 1.3: Propose a targeted patch
\hspace{0em}Based on your diagnosis, design ONE concrete, actionable fix for this sample:
\hspace{0em}- Target phase: Which phase of the process should be modified? (e.g., "Phase 2: Planning", "Phase 3: Execution")
\hspace{0em}- Constraint/Rule: State a concise, executable constraint or rule (e.g., "Before taking action X, verify condition Y", "When encountering situation Z, apply strategy W")
\hspace{0em}- Mechanism: How should this be implemented? (prompt engineering, constraint checking, search strategy change, etc.)
\hspace{0em}- Expected impact: What specific failure mode will this fix address? How will it reduce volatility or improve success rate?
\hspace{0em}Additionally, fill narrative\_summary: a standalone prose summary that complements the structured fields for Phase 2. Phase 2 receives the FULL Phase-1 JSON below (all schema fields) plus the original task text — not raw trajectory dumps.

\hspace{2em}=== OUTPUT FORMAT (STRICT JSON - NO MARKDOWN, NO CODE BLOCKS) ===

\hspace{0em}A single object with the same fields as one element of the historical per\_sample\_analysis array, PLUS narrative\_summary:
\hspace{0em}\texttt{\{}
\hspace{0em}\texttt{"task\_id": "string",}
\hspace{0em}\texttt{"num\_trajectories": int,}
\hspace{0em}\texttt{"score\_statistics": \{}
\hspace{0em}\texttt{"min": float, "max": float, "mean": float, "std": float, "gap": float}
\hspace{0em}\texttt{\},}
\hspace{0em}\texttt{"trajectory\_grouping": \{}
\hspace{0em}\texttt{"high\_performing\_indices": [], "low\_performing\_indices": [], "rationale": "string"}
\hspace{0em}\texttt{\},}
\hspace{0em}\texttt{"contrastive\_diagnosis": \{}
\hspace{0em}\texttt{"divergence\_points": ["string"],}
\hspace{0em}\texttt{"success\_factors": ["string"],}
\hspace{0em}\texttt{"failure\_modes": ["string"],}
\hspace{0em}\texttt{"volatility\_root\_cause": "string",}
\hspace{0em}\texttt{"difficulty\_root\_cause": "string"}
\hspace{0em}\texttt{\},}
\hspace{0em}\texttt{"candidate\_patch": \{}
\hspace{0em}\texttt{"target\_phase": "string",}
\hspace{0em}\texttt{"constraint\_rule": "string",}
\hspace{0em}\texttt{"implementation\_mechanism": "string",}
\hspace{0em}\texttt{"expected\_impact": "string"}
\hspace{0em}\texttt{\},}
\hspace{0em}\texttt{"narrative\_summary": "string"}
\hspace{0em}\texttt{\}}

\hspace{2em}=== CRITICAL REMINDERS ===

\hspace{4em}1. Be specific: Avoid vague statements like "trajectory fails due to poor reasoning". Instead: "trajectory fails at step 5 because it incorrectly assumes X when the constraint requires Y"

\hspace{4em}2. Use evidence: Ground every claim in concrete observations. Reference specific steps, actions, or outputs.

\hspace{4em}3. Think contrastively: Always compare high vs low trajectories. The DIFFERENCE is where the insight lies.

\hspace{4em}4. Focus on actionability: Every diagnosis should lead to a concrete, implementable fix. Avoid unfixable issues like "task is too hard".

\hspace{4em}5. Quantify when possible: Use numbers (frequencies, percentages, counts) to support claims about patterns.

\hspace{4em}6. Output pure JSON: No markdown code blocks, no extra text. Start with \{ and end with \}.

\hspace{4em}Begin your analysis now.

    \end{tcolorbox}
    \caption{Within-task reflection prompt in Skill-MAS evolution.}
    \label{fig:skill_mas_phase1}
\end{figure*}

\begin{figure*}[htbp]
    \centering
    \begin{tcolorbox}[
        enhanced,
        colframe=Salmon!90!Black,
        colback=Salmon!20,
        coltitle=white,
        fonttitle=\large\bfseries,
        title={Skill-MAS Evolution (Cross-Task Reflection)},
        halign title=left,
        fontupper=\footnotesize,
        boxrule=1pt,
        arc=3mm,
        boxsep=2pt,
        left=8pt,
        right=8pt,
        top=4pt,
        bottom=4pt
    ]

\hspace{0em}\textbf{System Prompt}

\hspace{2em}You are the diagnosis agent for Skill\_MAS Step 2 (Trajectory Reflection Synthesis). Your task is to synthesize cross-sample findings for Step3 optimization. For each task you receive the original problem text plus the COMPLETE Phase-1 structured JSON (score\_statistics, trajectory\_grouping, contrastive\_diagnosis, candidate\_patch, narrative\_summary, etc.). You do not receive raw per-agent trajectory dumps or mas\_code. Output strict JSON: cross\_sample\_synthesis and meta\_analysis only.

\hspace{0em}\textbf{User Prompt}

\hspace{2em}[Step2: Contrastive Trajectory Analysis — Phase 2 (Cross-sample only)]

\hspace{2em}=== SELECTION CONTEXT ===

\hspace{4em}Selected task\_ids (n=\{n\}): \{task\_id\_list\}

\hspace{4em}Input statistics:

\hspace{6em}- Character count: \{input\_char\_count\}

\hspace{6em}- Estimated tokens: \textasciitilde\{estimated\_input\_tokens\}

\hspace{4em}Selection rationale: These samples exhibit HIGH PRIORITY due to:

\hspace{6em}1. High cross-trajectory volatility: Large score variance across rollouts indicates unstable/inconsistent policy behavior

\hspace{6em}2. High intrinsic difficulty: Low average scores suggest systematic capability gaps

\hspace{2em}=== INPUT DATA ===

\hspace{4em}Phase 1 already analyzed each task's rollouts. Below, each block contains:

\hspace{6em}(1) the original problem / instruction text, and

\hspace{6em}(2) the COMPLETE Phase-1 structured JSON for that task — every field in the Phase-1 schema (task\_id, num\_trajectories, score\_statistics, trajectory\_grouping, contrastive\_diagnosis, candidate\_patch, narrative\_summary).

\hspace{4em}You do NOT have raw per-agent workflow dumps or mas\_code; use the structured Phase-1 JSON and task text only.

\hspace{4em}\{per\_task\_blocks\}

\hspace{2em}=== ANALYSIS FRAMEWORK ===

\hspace{4em}LEVEL 2: CROSS-SAMPLE SYNTHESIS (Global Pattern Recognition)

\hspace{4em}After Phase 1 analyzed each sample individually, synthesize global insights:

\hspace{4em}Step 2.1: Identify systematic weaknesses

\hspace{6em}- What failure modes appear across MULTIPLE samples? (frequency matters!)

\hspace{6em}- Are there common capability gaps? (e.g., "struggles with multi-step reasoning", "poor constraint checking")

\hspace{6em}- Are there common volatility sources? (e.g., "unstable at decision point X", "sensitive to prompt phrasing")

\hspace{6em}- Rank weaknesses by: (severity × frequency)

\hspace{4em}Step 2.2: Identify systematic strengths

\hspace{6em}- What does the policy consistently do well across samples?

\hspace{6em}- Are there patterns in successful trajectories that can be reinforced?

\hspace{6em}- What capabilities are reliable and can be leveraged?

\hspace{4em}Step 2.3: Prioritize fixes for Step 3

\hspace{6em}Based on the above, propose 3-5 prioritized fixes:

\hspace{6em}- Each fix should address a HIGH-IMPACT weakness (affects multiple samples or causes severe failures)

\hspace{6em}- Each fix should be ACTIONABLE (clear what to change and how)

\hspace{6em}- Rank by expected ROI: (impact × feasibility)

\hspace{6em}Format each fix as:

\hspace{6em}"[Priority X] <Fix description>: <Rationale> → Expected to improve <metric> on <affected samples>"

\vspace{4pt}
\hspace{2em}=== OUTPUT FORMAT (STRICT JSON - NO MARKDOWN, NO CODE BLOCKS) ===

\hspace{4em}\texttt{\{"cross\_sample\_synthesis": \{"systematic\_weaknesses": [\{"weakness": "string", "severity": "high|medium|low", "frequency": "string", "affected\_samples": [], "manifestation": "string"\}], "systematic\_strengths": [\{"strength": "string", "evidence": "string", "leverage\_opportunity": "string"\}], "prioritized\_fixes": [\{"priority": float, "fix\_description": "string", "rationale": "string", "implementation": "string", "expected\_impact": "string", "affected\_samples": int, "estimated\_effort": "low|medium|high"\}]\}, "meta\_analysis": \{"overall\_diagnosis": "string", "confidence\_level": "high|medium|low", "data\_quality\_notes": "string", "recommended\_next\_steps": ["string"]\}\}}

\hspace{2em}=== CRITICAL REMINDERS ===

\hspace{4em}1. Be specific: Tie weaknesses/strengths to task\_ids and concrete themes from the summaries when possible.

\hspace{4em}2. Use evidence: Ground claims in the Phase-1 structured outputs and task text — do not invent unseen trajectory detail.

\hspace{4em}3. Think globally: Patterns across samples drive prioritization.

\hspace{4em}4. Focus on actionability: prioritized\_fixes must be implementable in Step 3.

\hspace{4em}5. Quantify when possible: Use counts where summaries allow.

\hspace{4em}6. Output pure JSON: No markdown code blocks, no extra text. Start with \{ and end with \}.

\hspace{4em}Begin your synthesis now.

    \end{tcolorbox}
    \caption{Cross-task reflection prompt in Skill-MAS evolution.}
    \label{fig:skill_mas_phase2}
\end{figure*}

\begin{figure*}[htbp]
    \centering
    \begin{tcolorbox}[
        enhanced,
        colframe=Salmon!90!Black,
        colback=Salmon!20,
        coltitle=white,
        fonttitle=\large\bfseries,
        title={Skill-MAS Evolution (Skill Optimization)},
        halign title=left,
        fontupper=\footnotesize,
        boxrule=1pt,
        arc=3mm,
        boxsep=2pt,
        left=8pt,
        right=8pt,
        top=4pt,
        bottom=4pt
    ]

\hspace{0em}\textbf{System Prompt}
\hspace{0em}You are an expert author and optimizer for Skill-MAS three-stage SKILL.md files. Your task is to improve the current SKILL.md based on Step2 reflection evidence while preserving a valid 3-stage structure.

\hspace{0em}\textbf{User Prompt}
\hspace{0em}Evolution Context
\hspace{0em}- Round: \{round\_idx+1\} / \{total\_rounds\} (0-based round\_idx=\{round\_idx\})
\hspace{0em}- Target: Evolve a DOMAIN-AGNOSTIC, high-level Meta-Agent SKILL.md.
\hspace{0em}- Benchmark context (for understanding context only): \{bench\_hint\}

\hspace{2em}CURRENT SKILL.md
\hspace{0em}\{current\_skill\_md\}

\hspace{2em}STEP2 REFLECTION ANALYSIS (Your Core Evidence)
\hspace{0em}\{step2\_reflection\_summary\}

\hspace{2em}\textbf{YOUR MISSION: ARCHITECTURAL \& COGNITIVE EVOLUTION}
\hspace{0em}You are an elite AI Systems Architect. Your task is to evolve the provided SKILL.md. The goal is to elevate the Meta-Agent's capability to design sophisticated, robust, and generalizable Multi-Agent Systems (MAS).

\hspace{2em}\textbf{CRITICAL ABSTRACTION FIREWALL (MUST READ)}:

\hspace{4em}The Meta-Agent operates at the ARCHITECTURAL level, not the implementation level.
\hspace{0em}1. ABSOLUTELY NO domain-specific examples (e.g., DO NOT mention delivery, bookings, math formulas, or specific APIs).
\hspace{0em}2. ABSOLUTELY NO coding/syntax details (e.g., DO NOT mention Python, AST parsing, variable names, snake\_case, await, or state dictionaries).
\hspace{0em}3. ABSOLUTELY NO hardcoded heuristics (e.g., DO NOT say "if text contains 'and', split it").
\hspace{0em}Instead, "Actionable" means providing powerful COGNITIVE FRAMEWORKS and SYSTEM DESIGNS.

\hspace{2em}\textbf{PRUNING \& REFINEMENT (Before Adding)}

\hspace{4em}Before introducing new upgrades, review the CURRENT SKILL.md against the STEP2 REFLECTION ANALYSIS:
\hspace{0em}- IDENTIFY any existing guidance that the Step2 evidence suggests is ineffective, misleading, or counterproductive. Remove or rewrite it.
\hspace{0em}- CONDENSE overlapping or redundant points into a single, sharper formulation.
\hspace{0em}- CONSTRAINT: You may remove at most ONE existing element per stage section. When in doubt, keep it. Only delete when the Step2 evidence directly contradicts the guidance.
\hspace{0em}Then proceed to following STAGE 1-3 for new additions.

\hspace{2em}\textbf{OPTIMIZATION PROTOCOL BY STAGE}

\hspace{4em}Review the Step2 Reflection and extract the FUNDAMENTAL REASONING GAPS. Then, upgrade the three stages using the following high-level MAS concepts:

\hspace{4em}STAGE 1: TASK DECOMPOSITION MODULE (The "What")

\hspace{6em}Focus on elevating the planning paradigm.

\hspace{6em}- Look for gaps in handling uncertainty, implicit constraints, or complex dependencies.

\hspace{6em}- Substantive upgrades could include: "Constraint Verification Matrices", "Assumption Elicitation", "Milestone-based Evaluation", or identifying "Critical Paths" vs. "Optional Enhancements".

\hspace{6em}- Rule: Do not dictate how to calculate things; dictate how to structure the logical problem space.

\hspace{4em}STAGE 2: AGENT ENGINEERING MODULE (The "Who")

\hspace{6em}Focus on elevating agent autonomy and interaction contracts.

\hspace{6em}- Look for gaps in agent capability limits, hallucination, or poor instruction comprehension.

\hspace{6em}- Substantive upgrades could include: Designing "Standardized Input/Output Schemas (Contracts)", implementing "Self-Correction/Reflection Prompts" within individual agents, defining "Boundary Conditions" (when an agent should stop and ask for help), or establishing "System Prompts" vs "Task Prompts" isolation.

\hspace{6em}- Rule: Do not dictate specific task execution steps; dictate agent psychology and boundary definitions.

\hspace{4em}STAGE 3: WORKFLOW \& ORCHESTRATION MODULE (The "How")

\hspace{6em}Focus on elevating topological robustness and system resilience.

\hspace{6em}- Look for gaps in information loss between agents, infinite loops, or catastrophic pipeline failures.

\hspace{6em}- Substantive upgrades could include: Advanced Topologies (e.g., "Actor-Critic pairs for quality control", "Dynamic Routing based on confidence scores", "Hierarchical delegation"), "Context Compression" (preventing context window overflow during agent handoffs), or "Global vs. Local Memory Management".

\hspace{6em}- Rule: Do not dictate code debugging logic; dictate dataflow architecture and topological resilience patterns.

\hspace{2em}\textbf{EXECUTION \& OUTPUT CONSTRAINTS}

\hspace{4em}1. Evidence-Driven Abstraction: Every change must resolve a flaw found in Step2, but the solution MUST be abstracted into a universal systems-engineering principle.

\hspace{4em}2. Meaningful Depth: Do not just add adjectives. Add new sub-bullet points that introduce a concrete conceptual framework (e.g., instead of "Make dependencies clear", use "Build a Directed Acyclic Graph (DAG) mapping of logic state transitions").

\hspace{4em}3. Incremental evolution (hard limit): In this round, introduce at most one substantive conceptual upgrade per SKILL stage section (1, 2, 3 — each stage at most one focused improvement). Do not pile on multiple unrelated changes in a single pass.

\hspace{4em}4. Output Format: Produce ONLY the complete updated SKILL.md.

\hspace{4em}Format Requirements:

\hspace{6em}- MUST start directly with the YAML frontmatter (\texttt{---}).

\hspace{6em}- MUST preserve the exact same YAML keys and the exactly three-stage markdown structure (1, 2, 3).

\hspace{6em}- NO markdown code fences around the entire output.

\hspace{6em}- NO preamble, NO explanations, NO summary of changes. Output raw SKILL.md text only.

\hspace{4em}BEGIN YOUR EVOLUTION NOW.

    \end{tcolorbox}
    \caption{Skill optimization prompt for Skill-MAS evolution.}
    \label{fig:skill_mas_optimization}
\end{figure*}

\end{document}